\documentclass[12pt, draftclsnofoot, onecolumn]{IEEEtran}
\usepackage{cite}
\usepackage{amsmath,amssymb,amsfonts}
\usepackage{algorithm}  
\usepackage{algpseudocode}  
\usepackage{algpseudocode}
\usepackage{graphicx}
\usepackage{epstopdf}
\usepackage{textcomp}
\usepackage{xcolor}
\usepackage{bm}
\usepackage{setspace}

\ifCLASSINFOpdf
\else
\fi
\begin{document}
%
\title{Joint Beamforming Design and Power \\Splitting Optimization in IRS-Assisted SWIPT NOMA Networks}
%
%
%

\author{Zhendong~Li,~Wen~Chen,~\IEEEmembership{Senior~Member,~IEEE},~Qingqing~Wu,~\IEEEmembership{Member,~IEEE},~Kunlun~Wang,~\IEEEmembership{Member,~IEEE},~and~Jun~Li,~\IEEEmembership{Senior~Member,~IEEE}
	\thanks{Z. Li and W. Chen are with the Department of Electronic Engineering, Shanghai Jiao Tong University, Shanghai 200240, China (e-mail: lizhendong@sjtu.edu.cn; wenchen@sjtu.edu.cn).}
	\thanks{Q. Wu is with the State Key Laboratory of Internet of Things for Smart City, University of Macau, Macau, China (email: qingqingwu@um.edu.mo). }
	\thanks{K. Wang is with the School of Information Science and Technology, ShanghaiTech University, Shanghai 201210, China, and also with the School of Communication and Electronic Engineering, East China Normal University, Shanghai 200241, China (e-mail: wangkl2@shanghaitech.edu.cn).}
	\thanks{J. Li is with the School of Electronic and Optical Engineering, Nanjing University of Science Technology, Nanjing 210094, China (email: jun.li@njust.edu.cn). }
	\thanks{(\emph{Corresponding author: Wen Chen.})}}
\maketitle

\begin{abstract}
	This paper proposes a novel network framework of intelligent reflecting surface (IRS)-assisted simultaneous wireless information and power transfer (SWIPT) non-orthogonal multiple access (NOMA) networks, where IRS is used to enhance the NOMA performance and the wireless power transfer (WPT) efficiency of SWIPT. We formulate a problem of minimizing base station (BS) transmit power by jointly optimizing successive interference cancellation (SIC) decoding order, BS transmit beamforming vector, power splitting (PS) ratio and IRS phase shift while taking into account the quality-of-service (QoS) requirement and energy harvested threshold of each user. The formulated problem is non-convex optimization problem, which is difficult to solve it directly. Hence, a two-stage algorithm is proposed to solve the above-mentioned problem by applying semidefinite relaxation (SDR), Gaussian randomization and successive convex approximation (SCA). Specifically, after determining SIC decoding order by designing IRS phase shift in the first stage, we alternately optimize BS transmit beamforming vector, PS ratio, and IRS phase shift to minimize the BS transmit power. Numerical results validate the effectiveness of our proposed optimization algorithm in reducing BS transmit power compared to other baseline algorithms. Meanwhile, compared with non-IRS-assisted network, the IRS-assisted SWIPT NOMA network can decrease BS transmit power by 51.13\%.
\end{abstract}


%
\IEEEpeerreviewmaketitle

\section{Introduction}
%
%
%
%
\IEEEPARstart{W}{ith} the vigorous development of emerging services such as the Internet-of-Things (IoT) and mobile Internet, the surge in wireless devices poses unprecedented challenges to beyond fifth-generation/sixth-generation (B5G/6G) communication systems in terms of massive connectivity, spectrum efficiency, energy management, and deployment costs \cite{8910627,ni2020resource,6798744}. In order to meet the massive connectivity of future networks and the higher service requirements of users, non-orthogonal multiple access (NOMA) technology has triggered extensive discussions in academia and industry. Unlike conventional orthogonal multiple access (OMA) \cite{6995966}, NOMA can support multiple users to share the same resources, e.g., time, frequency, coding, etc., so it can support massive connectivity of users. Specifically, taking an instance of NOMA in the power domain, the base station (BS) uses the same resource blocks to serve multiple users, which can greatly improve the spectrum efficiency to meet the users' communication requirements \cite{liu2017non,7263349,8493528,7752784,8674826}. For power domain NOMA transmission in the downlink, superposition coding and successive interference cancellation (SIC) techniques are applied at the BS transmitter and at the users respectively \cite{liu2017non,8085125}. As such, users with stronger channel gains can remove co-channel interference caused by users with weaker channel gains before decoding \cite{7842433}.

In general, existing research on NOMA considers that the users' channel conditions differ greatly, i.e., there are one type of users near the BS, and the other type of users are at the edge of the BS. In this way, the BS will allocate more transmit power to users with poor channel conditions. The main reason for this consideration is that the large difference in channel conditions will greatly release the potential of NOMA \cite{7273963}. More specifically, if the difference in user channel conditions is small, the performance of NOMA is not much better than that of OMA. However, in practical scenarios, the difference in user channel conditions for the NOMA networks is not always very large. This is because the wireless channel is determined by the propagation environment, which is highly random and uncontrollable. Therefore, if the channel can be controlled and adjusted, the performance of NOMA will be greatly enhanced.

In addition, considering the requirements to provide continuous information transmission and energy transmission for large-scale low-power and energy-constrained IoT devices, simultaneous wireless information and power transfer (SWIPT) technology has attracted great attention recently. By applying SWIPT, users can obtain information and energy at the same time, which brings great convenience to the deployment of energy-constrained IoT devices \cite{7867826,8476597}. As one of the design schemes of practical SWIPT receivers, the power splitting (PS) scheme aims to split the signal received by the receiver into two different power streams with one part used to decode information and the other part used to harvest energy \cite{6805330}. Based on the PS scheme, \cite{6623062} proposed a novel integrated SWIPT receiver architecture to achieve miniaturization and energy saving of the receiver. However, for the conventional SWIPT system, the wireless power transfer (WPT) efficiency will decrease sharply as the distance increases due to severe propagation loss, which thus greatly limits the performance of the SWIPT system. If the channel conditions can be strengthened, the WPT efficiency and coverage will be improved.

Recently, intelligent reflecting surface (IRS) has been proposed as a promising cost-effective solution to control and adjust the wireless channel between transceivers \cite{8910627}, \cite{8811733,8930608,9048622}. Furthermore, it can greatly improve the spectrum efficiency, energy efficiency, and coverage of the networks, while reducing networking costs \cite{9136592,yuan2020reconfigurable}. Therefore, it has received widespread attention from academia and industry \cite{8466374} and has also been recognized as a key enabling for the future 6G ecosystem \cite{rajatheva2020white}. Specifically, IRS composed of a large number of passive reflecting elements can be easily deployed on indoor walls or buildings. It can adjust the amplitude and phase of the incident signal, and realize the reconstruction of wireless channel. Unlike conventional relays, IRS is a passive device, which only passively reflects the incident signal without signal processing, so it does not introduce additional noise \cite{gong2019towards,zhao2019survey}. Unlike multiple-input multiple-output (MIMO), the required hardware cost and power consumption are much lower. These have greatly promoted the application of IRS in B5G/6G networks. Based on these significant advantages of IRS, the momentum of the IRS-assisted SWIPT NOMA networks has been stimulated. Firstly, IRS reconstructs the wireless channel to make the difference in channel conditions among users, thereby enhancing the performance of NOMA. Besides, it can improve the WPT efficiency of SWIPT system, and expand the network coverage. In short, IRS-enhanced wireless networks can meet the challenges of future B5G/6G networks in terms of massive connectivity, spectrum efficiency, energy management, and cost, etc..

The application of IRS-assisted wireless networks in different scenarios and different technologies has continued to emerge, e.g., IRS-assisted MIMO \cite{9117136,9159923,9201173}, IRS-assisted massive MIMO \cite{9355404}, IRS-assisted mobile edge computing \cite{9133107}, IRS-assisted unmanned aerial vehicle (UAV) communication \cite{9234511,8959174}, IRS-assisted physical layer security \cite{9159923,9201173,8743496,9248012,9390203} and robust beamforming design in IRS-aided MISO communications \cite{9180053}, etc.. In addition, many scholars are currently committed to the research of IRS-enhanced NOMA transmission \cite{fu2019reconfigurable,8970580,9167258,9203956,9140006,9240028,ni2020resource} and the research of IRS-assisted SWIPT technology \cite{8941080,9133435,9110849}. For the research of IRS-enhanced NOMA, Fu \emph{et al.} jointly optimized BS beamforming vector and IRS phase shift in the NOMA network to minimize the total transmit power \cite{fu2019reconfigurable}. In \cite{9167258}, Zuo \emph{et al.} considered the NOMA network in the single-input single-output (SISO) scenario, with the goal of maximizing system throughput, jointly optimizing the decoding order, channel selection and the IRS phase shift matrix. In \cite{ni2020resource}, Ni \emph{et al.} proposed a new framework for resource allocation in multi-cell IRS-assisted NOMA networks and maximized achievable sum-rate. In \cite{9240028}, Zhu \emph{et al.} proposed an IRS-assisted downlink energy-efficient multiple-input single-output (MISO) transmission scheme, which greatly reduces the transmit power by optimizing the beamforming vector and the IRS phase shifs. Meanwhile, some progress has been made in the research on IRS-assisted SWIPT. In the IRS-assisted SWIPT networks, Wu \emph{et al.} jointly optimized the active and passive beamforming vector to increase the weighted received power of information user (IU) and energy user (EU) \cite{8941080}. Meanwhile, a novel algorithm was adopted in \cite{9133435} to solve the problem of maximizing the weighted received power of IU and EU under the condition of satisfying all users' QoS. Pan \emph{et al.} optimized the weighted sum-rate in IRS-assisted MIMO SWIPT networks \cite{9110849}.

Although there have been many researches on IRS-assisted wireless communication networks. However, the B5G/6G network will be a more complex and changeable network. The massive connectivity, spectrum efficiency, energy management and deployment cost of the network will greatly stimulate our motivation to integrate IRS with NOMA and SWIPT in order to better satisfy the business requirements of users in IoT networks. As far as we know, the current research on IRS-assisted SWIPT NOMA networks is still in its infancy. In this paper, under the constraints of meeting the users' QoS requirements and energy harvested thresholds, we minimize the BS transmit power by jointly optimizing the SIC decoding order, BS transmit beamforming vector, received PS ratio, and IRS phase shift. This problem is challenging mainly because the changes of wireless channel make the users' decoding order more complicated, and the BS transmit beamforming vector, PS ratio, and IRS phase shift are highly coupled. Therefore, it is necessary to design an effective algorithm for the IRS-assisted SWIPT NOMA networks to minimize the BS transmit power. 

Based on the above, the main contributions of this paper can be summarized as follows:
\begin{itemize}
\item We propose an IRS-assisted SWIPT NOMA network framework, in which multiple users can share the same resource blocks at the same time, and users can obtain energy while receiving information. In addition, IRS can adjust the channel to improve NOMA performance and WPT efficiency of SWIPT. We formulate the BS transmit power minimization problem for joint optimization of the SIC decoding order, BS transmit beamforming vector, PS ratio, and IRS phase shift. Since the optimization variables are highly coupled, solving this problem is challenging.

\item In order to solve the optimization problem, we divide the problem into two stages. Specifically, in the first stage, an SIC decoding order determination algorithm is proposed based on the maximum combined channel gain. In the second stage, we divide the problem into three sub-problems according to the decoding order obtained in the first stage. Firstly, given PS ratio and IRS phase shift, the BS transmit beamforming vector is optimized by applying semidefinite relaxation (SDR) and successive convex optimization (SCA). For the last two feasibility-check problems of PS ratio and IRS phase shift, we efficiently solve them by applying SDR, SCA and Gaussian randomization. Finally, the three sub-problems are iterated alternately until convergence.

\item Through numerical simulation, we verified the effectiveness of the proposed joint SIC decoding order, BS transmit beamforming vector, PS ratio, and IRS phase shift optimization algorithm (JDBPR) compared with the baseline algorithm, i.e., it can significantly decrease the BS transmit power. For the IRS-assisted NOMA SWIPT networks, the BS transmit power is significantly lower than the networks without IRS assistance. Meanwhile, the more reflecting elements of the IRS, the smaller the required BS transmit power, which also means that we can reduce the BS transmit power by increasing the number of IRS elements.
\end{itemize}
%
%

The remainder of this paper is organized as follows. Section II elaborates the system model and optimization problem formulation for the IRS-assisted SWIPT NOMA networks. Section III presents the proposed two-stage optimization algorithm for the formulated optimization problem. In Section IV, numerical results demonstrate that our algorithm has good convergence and effectiveness. Finally, the conclusion is given in Section V.

\textit{Notations:} Scalars are denoted by lower-case letters, while vectors and matrices are represented by bold lower-case letters and bold upper-case letters, respectively. $\left| {x} \right|$ denotes the absolute value of a complex-valued scalar $x$, and $\left\| {\bf{x}} \right\|$ denotes the Euclidean norm of a complex-valued vector $\bf{x}$. $diag(\bf{x})$ denotes a diagonal matrix whose diagonal elements are the corresponding elements in vector $\bf{x}$. For a square matrix $\bf{X}$, $\rm{Tr(\bf{X})}$, $\rm{Rank(\bf{X})}$, ${\bf{X}}^H$ and ${\bf{X}}_{m,n}$ denote its trace, rank, conjugate transpose and $m,n$-th entry, respectively, while ${\bf{X}} \succeq 0$ represents that $\bf{X}$ is a positive semidefinite matrix. Similarly, for a general matrix $\bf{A}$, $\rm{Tr(\bf{A})}$, $\rm{Rank(\bf{A})}$, ${\bf{A}}^H$ and ${\bf{A}}_{m,n}$ also denote its trace, rank, conjugate transpose and $m,n$-th entry, respectively. In addition, ${\mathbb{C}^{M \times N}}$ denotes the space of ${M \times N}$ complex matrices. ${\bf{I}}_N$ denotes an dentity matrix of size ${N \times N}$. $j$ denotes the imaginary unit, i.e., $j^2=-1$. Finally, the distribution of a circularly symmetric complex Gaussian (CSCG) random vector with mean $\mu$ and covariance matrix $\bf{C}$ is denoted by $ {\cal C}{\cal N}\left( {\mu,\bf{C}} \right)$, and $\sim$ stands for `distributed as'.
\section{System model and problem formulation}
\subsection{System Model}
In this paper, we consider the downlink transmission in an IRS-assisted SWIPT NOMA network consisting of one BS, one IRS and $K$ users. The set of users is denoted by ${\cal K} = \left\{ {1,2,...,K} \right\}$. It can be assumed that the BS is equipped with $N > 1$ uniform linear array (ULA) antennas and each user is equipped with one antenna. The IRS is equipped with $M$ ULA reflecting elements\footnote{The proposed optimization algorithm when the IRS is equipped with ULA can be well extended to the setup where the IRS is equipped with uniform planar array (UPA).}, denoted by ${\cal M} = \left\{ {1,2,...,M} \right\}$. A smart controller is equipped at the IRS to coordinate its switching between two working modes, namely the receiving mode for channel estimation and the reflection mode for signal transmission \cite{6231145}. Since IRS elements are passively reflective, they are passive devices. We consider the use of time-division duplexing (TDD) protocol for uplink and downlink communication. Combined with the reciprocity of the channel, the downlink channel state information (CSI) can be obtained according to the uplink channel estimation. Thus, we assume the CSI of all channels is perfectly known at the BS\footnote{It is worth noting that although this paper is the design of the optimization algorithm with the assumption that perfect CSI is available at the BS, the framework and process of the proposed optimization algorithm can still be applied to robust optimization algorithm design in the case of imperfect CSI.}. Meanwhile, we assume that all channels are quasi-static flat-fading.

\begin{figure}
	\centerline{\includegraphics[width=7cm]{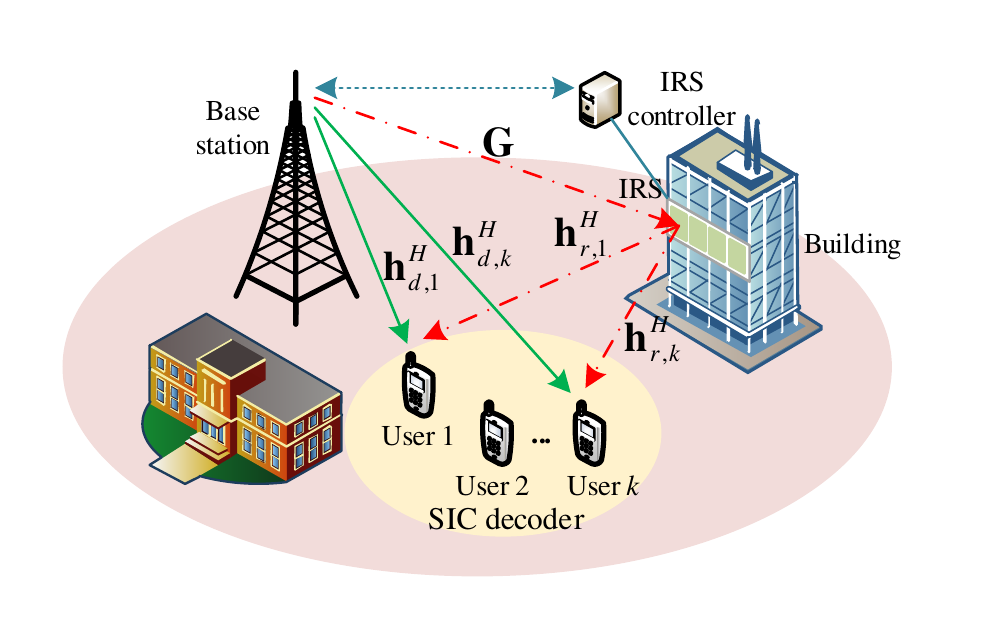}}
	\caption{The IRS-assisted SWIPT NOMA networks.}
	\label{Fig1}
\end{figure}

The channel gains from the BS to IRS, from the IRS to $k$-th user, and from the BS to $k$-user are respectively represented by ${\bf{G}} \in {\mathbb{C}^{M \times N}}$, ${\bf{h}}_{r,k}^H \in {\mathbb{C}^{1 \times M}}$ and ${\bf{h}}_{d,k}^H \in {\mathbb{C}^{1 \times N}}$, $\forall k \in {\cal K}$. Let ${\bf{\Theta }} = {\rm{diag}}\left( {{\beta _1}{e^{j{\theta _1}}},{\beta _2}{e^{j{\theta _2}}},...,{\beta _M}{e^{j{\theta _M}}}} \right) \in {\mathbb{C}^{M \times M}}$ denote the reflection coefficients matrix of the IRS, where ${\beta _m} \in \left[ {0,1} \right]$ and ${\theta _m} \in \left[ {0,2\pi } \right]$ denote the amplitude reflection coefficient and phase shift of the $m$-th reflecting element, respectively\footnote{It is worth noting that there are IRS amplitude and phase shift models that are closer to the practical system. In order to characterize the basic performance limit of IRS, we assume that the phase shift varies continuously from 0 to 2$\pi$. In practice systems, we usually choose from a discrete value from 0 to 2$\pi$. Discrete phase shift optimization research will be discussed in the future work. In addition, the amplitude of the IRS can also vary continuously from 0 to 1. In practice, in order to maximize the power of the reflected signal, we usually set $\beta_m=1$.}. Due to the severe path loss, signals reflected by the IRS twice or more are negligible and can be ignored. In the deployment of actual scenarios, we usually consider that the reflecting element of the IRS is designed to maximize the reflected signal, thus ${\beta _m}{\rm{ = 1,}} \forall m \in {\cal M}$. The wireless channel can be divided into three parts, namely, BS-IRS channel, IRS-user channel, BS-user channel. Although the channel between the BS and users may be blocked, the wireless channel still has a lot of scattering, thus we model the BS-user channel as Rayleigh fading and denote it as ${\bf{g}}_{d,k}^H \in {{\mathbb{C}}^{1 \times N}}$. We assume that each element of ${\bf{g}}_{d,k}^H$ is independent and identically distributed (i.i.d.) CSCG random variable with zero mean and unit variance. Therefore, the channel gain of the BS-user can be expressed as
\begin{equation}
{\bf{h}}_{d,k}^H = \sqrt {{C_0}{{\left( {\frac{{{d_{d,k}}}}{{{D_0}}}} \right)}^{ - {\alpha}_1 }}} {\bf{g}}_{d,k}^H,
\end{equation}
where $C_0$ represents the path loss when the reference distance ${D_0} = 1\left( {\rm{m}} \right)$, ${{d_{d,k}}}$ is the distance from the BS to the $k$-th user, and ${\alpha}_1 $ represents the path loss exponent.

In addition, for the BS-IRS channel and IRS-user channel, there are LoS components, thus we model them as Rician fading. The BS-IRS channel can be denoted by
\begin{equation}
{\bf{\bar G}} = \sqrt {\frac{\kappa }{{1 + \kappa }}} {{\bf{G}}^{{\rm{LoS}}}} + \sqrt {\frac{1}{{1 + \kappa }}} {{\bf{G}}^{{\rm{NLoS}}}},
\end{equation}
where $\kappa$ is the Rician factor, ${{\bf{G}}^{{\rm{LoS}}}} \in {\mathbb{C}^{M \times N}}$ and ${{\bf{G}}^{{\rm{NLoS}}}} \in {\mathbb{C}^{M \times N}}$ are the line-of-sight (LoS) component and non-line-of-sight (NLoS) component, respectively. Each element of ${{\bf{G}}^{{\rm{NLoS}}}}$ is i.i.d. CSCG random variable with zero mean and unit variance. Similarly, the IRS-user channel can be expressed as
\begin{equation}
{\bf{\bar h}}_{r,k}^H = \sqrt {\frac{\vartheta }{{1 + \vartheta }}} {\bf{h}}_{r,k}^{{\rm{LoS}}} + \sqrt {\frac{1}{{1 + \vartheta }}} {\bf{h}}_{r,k}^{{\rm{NLoS}}},
\end{equation}
where $\vartheta$ is the Rician factor, ${\bf{h}}_{r,k}^{{\rm{LoS}}} \in {\mathbb{C}^{1 \times M}}$ and ${\bf{h}}_{r,k}^{{\rm{NLoS}}} \in {\mathbb{C}^{1 \times M}}$ are the LoS component and NLoS component, respectively. Each element of ${\bf{h}}_{r,k}^{{\rm{NLoS}}}$ is i.i.d. CSCG random variable with zero mean and unit variance.

The LoS component is represented by the array response of ULA. The array response of $N$ elements ULA can be expressed as
\begin{equation}
{{\bf{a}}_N}\left( \theta  \right) = \left[ {1,{e^{ - j2\pi \frac{d}{\lambda }\sin \theta }},...,{e^{ - j2\pi \left( {N - 1} \right)\frac{d}{\lambda }\sin \theta }}} \right],
\end{equation}
where $\theta$ represents the angle of arrival (AoA) or angle of departure (AoD) of the signal. Therefore, the LoS component ${\bf{G}}^{{\rm{LoS}}}$ can be given by
\begin{equation}
{{\bf{G}}^{{\rm{LoS}}}} = {\bf{a}}_M^H\left( {{\theta _{{\rm{AoA,1}}}}} \right){{\bf{a}}_N}\left( {{\theta _{{\rm{AoD,1}}}}} \right),
\end{equation}
where ${{\theta _{{\rm{AoA,1}}}}}$ is the AoA to the ULA at the IRS, and
${{\theta _{{\rm{AoD,1}}}}}$ is the AoD from the ULA at the BS. Similarly, the LoS component ${\bf{h}}_{r,k}^{{\rm{LoS}}}$ can be expressed as
\begin{equation}
{\bf{h}}_{r,k}^{{\rm{LoS}}} = {{\bf{a}}_M}\left( {{\theta _{{\rm{AoD,2}}}}} \right),
\end{equation}
where ${{\theta _{{\rm{AoD,2}}}}}$ is the AoD from the ULA at the IRS.

Therefore, the channel gain of BS-IRS and IRS-user can be expressed as
\begin{equation}
{\bf{G}} = \sqrt {{C_0}{{\left( {\frac{{{d_{d,r}}}}{{{D_0}}}} \right)}^{ - {\alpha}_2 }}} {\bf{\bar G}},
\end{equation}
and
\begin{equation}
{\bf{h}}_{r,k}^H = \sqrt {{C_0}{{\left( {\frac{{{d_{r,k}}}}{{{D_0}}}} \right)}^{ - {\alpha}_3}}} {\bf{\bar h}}_{r,k}^H,
\end{equation}
where ${{d_{d,r}}}$ and ${{d_{r,k}}}$ represent the distance from the BS to the IRS and the distance from the IRS to the $k$-th user, respectively. ${\alpha}_2$ and ${\alpha}_3$ respectively represent the path loss exponent from BS to IRS and IRS to the $k$-th user. 

In this paper, we assume linear transmit precoding at BS, where each user is assigned with one dedicated information beam. Therefore, the complex baseband transmitted signal at BS can be expressed as
\begin{equation}
{\bf{x}} = \sum\limits_{k = 1}^K {{{\bf{w}}_k}{s_k}} ,\forall k \in {\cal K},
\end{equation}
where ${s_k}$ denotes the transmission data symbol for the $k$-th user, and ${{\bf{w}}_k} \in {\mathbb{C}^{N \times 1}}$ represents the corresponding beamforming vector. We assume that ${s_k}$ is i.i.d. CSCG random variables with zero mean and unit variance, denoted by ${s_k} \sim {\cal C}{\cal N}\left( {0,1} \right),\forall k \in {\cal K}$.

The received signal at the $k$-th user from both the BS-user link and BS-IRS-user link can be expressed as
\begin{equation}
{y_k} = \left( {{\bf{h}}_{r,k}^H{\bf{\Theta G}} + {\bf{h}}_{d,k}^H} \right)\sum\limits_{j = 1}^K {{{\bf{w}}_j}{s_j}}  + {n_k},\forall k \in {\cal K},
\end{equation}
where ${n_k} \sim {\cal C}{\cal N}\left( {0,\sigma _k^2} \right)$ denotes the antenna noise at the $k$-th user.

In addition, we consider that each user applies PS scheme to coordinate the process of information decoding and energy harvested from the received signal \cite{6805330}. The PS receiver architecture is shown in Fig. 2. The received signal at each user is split to the information decoder (ID) and the energy harvester (EH) by a power splitter. For the $k$-th user, it divides ${\rho _k}\left( {0 \le {\rho _k} \le 1} \right)$ portion of the signal power to the ID, remaining $1 - {\rho _k}$ portion of the signal power to the EH. Therefore, the signal split to the ID for $k$-th user can be given by
\begin{equation}
y_k^{{\rm{ID}}} = \sqrt {{\rho _k}} \left( {\left( {{\bf{h}}_{r,k}^H{\bf{\Theta G}} + {\bf{h}}_{d,k}^H} \right)\sum\limits_{j = 1}^K {{{\bf{w}}_j}{s_j}}  + {n_k}} \right) + {z_k},\forall k \in {\cal K},
\end{equation}
where ${z_k} \sim {\cal C}{\cal N}\left( {0,\delta _k^2} \right)$ is the additional noise introduced by the ID at the $k$-th user.

\begin{figure}
	\centerline{\includegraphics[width=8cm]{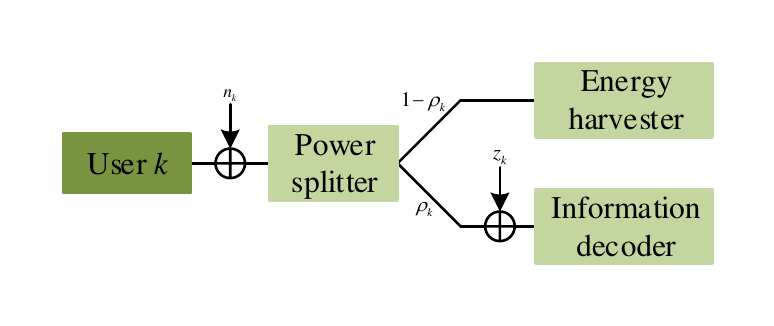}}
	\caption{The PS receiver architecture.}
	\label{Fig2}
\end{figure}

Since we consider NOMA transmission, the SIC decoding order is very important, which is determined by the channel conditions. Unlike the general NOMA communication system, the combination of IRS and NOMA makes the channel more complicated, because the channel gain may change due to the change of the IRS phase shift matrix. Let $s\left( k \right)$ denote the decoding order for the $k$-th user. Then, $s\left( k \right) = i$ denotes that the $k$-th user is the $i$-th signal to be decoded at the receiver. Accordingly, the signal to interference-plus-noise-ratio (SINR) of the $k$-th user can be expressed as
\begin{equation}
{\rm{SIN}}{{\rm{R}}_k} = \frac{{{\rho _k}{{\left| {\left( {{\bf{h}}_{r,k}^H{\bf{\Theta G}} + {\bf{h}}_{d,k}^H} \right){{\bf{w}}_k}} \right|}^2}}}{{{\rho _k}\sum\limits_{s\left( j \right) > s\left( k \right)} {{{\left| {\left( {{\bf{h}}_{r,k}^H{\bf{\Theta G}} + {\bf{h}}_{d,k}^H} \right){{\bf{w}}_j}} \right|}^2}}  + {\rho _k}\sigma _k^2 + \delta _k^2}}.
\end{equation}
It is assumed that $s\left( k \right) \le s\left( {\bar k} \right)$, the SINR of the $\bar k$-th user decoding information signal of the $k$-th user can be given by
\begin{equation}
{\rm{SIN}}{{\rm{R}}_{\bar k \to k}} = \frac{{{\rho _{\bar k}}{{\left| {\left( {{\bf{h}}_{r,\bar k}^H{\bf{\Theta G}} \!+\! {\bf{h}}_{d,\bar k}^H} \right){{\bf{w}}_k}} \right|}^2}}}{{{\rho _{\bar k}}\sum\limits_{s\left( j \right) > s\left( k \right)} {{{\left| {\left( {{\bf{h}}_{r,\bar k}^H{\bf{\Theta G}} \!+\! {\bf{h}}_{d,\bar k}^H} \right){{\bf{w}}_j}} \right|}^2}} \! + \!{\rho _{\bar k}}\sigma _{\bar k}^2 \!+\! \delta _{\bar k}^2}}.
\end{equation}
In order to ensure that the $\bar k$-th user can decode the information of the $k$-th user with the decoding order $s\left( k \right) \le s\left( {\bar k} \right)$, the SIC decoding condition $ {\rm{SIN}}{{\rm{R}}_k}\le{\rm{SIN}}{{\rm{R}}_{\bar k \to k}}$ should be satisfied \cite{8170332}. For example, supposing the decoding order of the three users is $s\left( k \right) = i,i = 1,2,3$. Therefore, the SIC decoding conditions at user 2 and user 3 should satisfy the following conditions: ${\rm{SIN}}{{\rm{R}}_{2 \to 1}} \ge {\rm{SIN}}{{\rm{R}}_1}$, ${\rm{SIN}}{{\rm{R}}_{3 \to 1}} \ge {\rm{SIN}}{{\rm{R}}_1}$ and ${\rm{SIN}}{{\rm{R}}_{3 \to 2}} \ge {\rm{SIN}}{{\rm{R}}_2}$. 

In addition, the signal split to the EH for the $k$-th user can be expressed as 
\begin{equation}
y_k^{{\rm{EH}}} = \sqrt {1 - {\rho _k}} \left( {\left( {{\bf{h}}_{r,k}^H{\bf{\Theta G}} + {\bf{h}}_{d,k}^H} \right)\sum\limits_{j = 1}^K {{{\bf{w}}_j}{s_j}}  + {n_k}} \right),
\end{equation}
Then, the harvested power by the EH for the $k$-th user can be given by
\begin{equation}
{E_k} = {\eta _k}\left( {1 - {\rho _k}} \right)\left( {\sum\limits_{j = 1}^K {{{\left| {\left( {{\bf{h}}_{r,k}^H{\bf{\Theta G}} + {\bf{h}}_{d,k}^H} \right){{\bf{w}}_j}} \right|}^2}}  + \sigma _k^2} \right),
\end{equation}
where ${\eta _k} \in \left( {0,1} \right]$ denotes the power conversion efficiency at EH of the $k$-th user. In this paper, we consider the normalized time, then the harvested power is the harvested energy.
\subsection{Problem Formulation for the IRS-assisted SWIPT NOMA Networks}

In this paper, we aim to minimize the BS transmit power by jointly designing SIC decoding order, BS transmit beamforming vector, received PS ratio and IRS phase shift matrix. Accordingly, the problem can be formulated as follows,
\begin{subequations}
	\begin{align}
	\left( {{\textrm{P1}}} \right){\rm{~~~~~}}&\mathop {\min }\limits_{\left\{ {{{\bf{w}}_k},{\theta _m},{\rho _k},s\left( k \right)} \right\}} {\rm{  }}\sum\limits_{k = 1}^K {{{\left\| {{{\bf{w}}_k}} \right\|}^2},} \\
	\rm{s.t.}\qquad &\frac{{{\rho _k}{{\left| {\left( {{\bf{h}}_{r,k}^H{\bf{\Theta G}}\! +\! {\bf{h}}_{d,k}^H} \right){{\bf{w}}_k}} \right|}^2}}}{{{\rho _k}\!\!\sum\limits_{s\left( j \right) > s\left( k \right)} \!{{{\left| {\left( {{\bf{h}}_{r,k}^H{\bf{\Theta G}}\!+\! {\bf{h}}_{d,k}^H} \right){{\bf{w}}_j}} \right|}^2}}  \!+\! {\rho _k}\sigma _k^2 \!+\! \delta _k^2}} \!\ge\! {\gamma _k},\\
	&{{\rm{SIN}}{{\rm{R}}_k}\le\rm{SIN}}{{\rm{R}}_{\bar k \to k}},~{\rm{if}}~{\rm{ }} s\left( k \right)\le s\left( {\bar k} \right),\\
	&{\eta _k}\left( {1 \!-\! {\rho _k}} \right)\!\left( {\sum\limits_{j = 1}^K \!{{{\left| {\left( {{\bf{h}}_{r,k}^H{\bf{\Theta G}} \!+\! {\bf{h}}_{d,k}^H} \right){{\bf{w}}_j}} \right|}^2}}\!  +\! \sigma _k^2} \right)\!\! \ge {e_k},\\
	&{\rm{0}} \le {\rho _k} \le1,\\
	&{\rm{0}} \le {\theta _m} \le 2\pi ,\\
	&s\left( k \right) \in \Omega ,
	\end{align}
\end{subequations}
where constraint (16b) guarantees the QoS requirement of the $k$-th user with the SINR threshold ${\gamma _k}$. Meanwhile, constraint (16c) ensures SIC decoding conditions. In addition, constraint (16d) requires that the energy harvested of the $k$-th user needs to reach a threshold ${e_k}$. Considering that each user should have non-zero SINR threshold and energy harvested threshold, i.e., ${\gamma _k} > 0$ and  ${e_k}>0$, the received PS ratio of the $k$-th user should satisfy constraint (16e). Constraint (16f) is the condition that IRS phase shift should meet. $\Omega $ in constraint (16g) is the combination set of all possible SIC decoding orders.

It can be seen that the problem (P1) is a non-convex optimization problem due to the following reasons. Firstly, the BS transmit beamforming, received PS ratio and IRS phase shift matrix are highly coupled. In addition, the phase shift is expressed in exponential form. Finally, the SIC decoding orders are determined by the IRS phase shift, thus the SIC decoding order and the IRS phase shift are also related. Therefore, it is challenging to directly solve this problem.

\section{The proposed two-stage optimization algorithm for the IRS-assisted SWIPT NOMA networks}
In this section, we propose a two-stage optimization algorithm to solve the problem (P1). The problem (P1) is decoupled into two stages. Firstly, an SIC decoding order determination algorithm based on combined channel gain is proposed. Then, for the given SIC decoding order, the beamforming vector, PS ratio and IRS phase shift matrix are alternately optimized by applying SDR, SCA and Gaussian randomization. 
\subsection{SIC Decoding Order Determination Algorithm}
In this subsection, the SIC decoding order determination algorithm based on the combined channel gains is proposed. Since this paper considers NOMA, the SIC decoding order is a factor that must be considered. In fact, the SIC decoding order is determined by the channel gain from the BS to each user. Due to the addition of IRS, the SIC decoding order is not only determined by the direct channel gain from the BS to the ground user, but by the combined channel gain from the BS to each user. The change of the IRS phase shift matrix will influence the combined channel gain. Since the same phase shift is different for all users, the combined channel gain of different users cannot be maximized at the same time. Therefore, we maximize sum of the combined channel gain from the BS to all users by optimizing the IRS phase shift. We can sort the combined channel gains of all users to determine the SIC decoding order.  The problem of maximizing the sum of the combined channel gains of all users can be expressed as 
\begin{subequations}
	\begin{align}
	\left( {{\textrm{P2}}} \right){\rm{~~~~}}&\mathop {{\rm{max}}}\limits_{\bf{\Theta }} {\rm{  }}\sum\limits_{k = 1}^K {{{\left\| {{\bf{h}}_{r,k}^H{\bf{\Theta G}} + {\bf{h}}_{d,k}^H} \right\|}^2}},  \\
	\rm{s.t.}~~~~~&{\rm{0}} \le {\theta _m} \le 2\pi,
	\end{align}
\end{subequations}
where ${\bf{\Theta }} = {\rm{diag}}\left( {{e^{j{\theta _1}}},{e^{j{\theta _2}}},...,{e^{j{\theta _M}}}} \right)$. Let ${u_m} = {e^{j{\theta _m}}},\forall m \in {\cal M}$, ${\bf{u}} = {\left[ {{u_1},{u_2},...,{u_M}} \right]^H} \in {\mathbb{C}^{M \times 1}}$. Then the constraint on ${\theta _m}$ is equivalent to $\left| {{u_m}} \right| = 1,\forall m \in {\cal M}$. Let ${{\bf{a}}_k} = {\rm{diag}}\left( {{\bf{h}}_{r,k}^H} \right){\bf{G}} \in {\mathbb{C}^{M \times N}}$, then ${\left\| {{\bf{h}}_{r,k}^H{\bf{\Theta G}} + {\bf{h}}_{d,k}^H} \right\|^2}$ can be written as ${\left\| {{{\bf{u}}^H}{{\bf{a}}_k} + {\bf{h}}_{d,k}^H} \right\|^2}$. We introduce auxiliary variables as follows,
\begin{equation}
{{\bf{R}}_k} = \left[ {\begin{array}{*{20}{c}}
{{{\bf{a}}_k}{\bf{a}}_k^H}&{{{\bf{a}}_k}{{\bf{h}}_{d,k}}}\\
{{\bf{h}}_{d,k}^H{\bf{a}}_k^H}&0
\end{array}} \right],{\bf{\bar u}}{\rm{ = }}\left[ {\begin{array}{*{20}{c}}
{\bf{u}}\\
1
\end{array}} \right].
\end{equation}
Therefore, ${\left\| {{{\bf{u}}^H}{{\bf{a}}_k} + {\bf{h}}_{d,k}^H} \right\|^2}$ can be further expressed as ${{\bf{\bar u}}^H}{{\bf{R}}_k}{\bf{\bar u}} + {\left\| {{\bf{h}}_{d,k}^H} \right\|^{\rm{2}}}$. Since ${{\bf{\bar u}}^H}{{\bf{R}}_k}{\bf{\bar u}} = {\rm{Tr}}\left( {{{\bf{R}}_k}{\bf{\bar u}}{{{\bf{\bar u}}}^H}} \right)$, we define ${\bf{\bar U}} = {\bf{\bar u}}{{\bf{\bar u}}^H}$, where ${\bf{\bar U}} \succeq 0$ and ${\rm{Rank}}\left( {\bf{\bar U}} \right) = 1$. Since the rank-one constraint is non-convex, we use SDR to relax this constraint firstly, and the problem (P2) can be transformed into
\begin{subequations}
	\begin{align}
	\left( {{\textrm{P2.1}}} \right){\rm{~~~~}}&\mathop {{\rm{max}}}\limits_{\bf{\bar U}} {\rm{  }}\sum\limits_{k = 1}^K {\left( {{\rm{Tr}}\left( {{{\bf{R}}_k}{\bf{\bar U}}} \right){\rm{ + }}{{\left\| {{\bf{h}}_{d,k}^H} \right\|}^{\rm{2}}}} \right)}, \\
	\rm{s.t.}\qquad &{{\bf{\bar U}}_{m,m}} = 1,m = 1,2,...,M + 1,\\
	&{\bf{\bar U}} \succeq 0.
	\end{align}
\end{subequations}
The above problem (P2.1) is a standard semidefinite programming (SDP) problem, which can be solved by using CVX toolbox \cite{grant2014cvx}. The problem (P2.1) is equivalent to the problem (P2) if and only if the optimal solution ${\bf{\bar U}}^*$ of the problem (P2.1) is a rank-one matrix. However, under normal circumstances, the problem (P2.1) generally does not produce a rank-one solution, i.e., ${\rm{Rank}}\left( {\bf{\bar U}} \right) \ne 1$. The optimal solution of problem (P2.1) is only the upper bound of problem (P2), so it is necessary to reconstruct the high-rank solution obtained from problem (P2.1) into a rank-one solution. In this paper, we use Gaussian randomization to reduce the rank of high-rank solution. Since ${\rm{Rank}}\left( {\bf{\bar U}} \right) \ne 1$, the eigenvalue decomposition of ${\bf{\bar U}}$ can be expressed as
\begin{equation}
{\bf{\bar U = V\Sigma }}{{\bf{V}}^H},
\end{equation}
where ${\bf{V}} = \left[ {{e_1},{e_2},...,{e_{M + 1}}} \right]$ is the identity matrix of the eigenvector, and ${\bf{\Sigma }}{\rm{ = diag}}\left( {{\lambda _{\rm{1}}},{\lambda _2},...,{\lambda _{M + 1}}} \right)$ is the diagonal matrix of the eigenvalue. Next, we generate two independent zero-mean normal distribution random vectors ${\bf{\alpha }} \in {\mathbb{R}^{\left( {M + 1} \right) \times 1}}$, ${\bf{\beta }} \in {\mathbb{R}^{\left( {M + 1} \right) \times 1}}$ and covariance matrix $\frac{1}{2}{{\bf{I}}_{M + 1}}$. Let $T$ be the maximum generation of candidate random variables. The Gaussian random vector of the $t$-th generation can be expressed as
\begin{equation}
{{\bf{r}}_t} = {\bf{\alpha }}_t + {\bf{\beta }}_t\sqrt { - 1} ,t = 1,2,...,T,
\end{equation}
where $\bf{\alpha }_t$ and $\bf{\beta }_t$ are the $t$-th generation random vectors. Based on the obtained Gaussian random vector ${{\bf{r}}_t} \sim {\cal C}{\cal N}\left( {{\bf{0}},{{\bf{I}}_{M + 1}}} \right)$, we can obtain the suboptimal solution of the problem (P2.1), which can be expressed as
\begin{equation}
{{\bf{\bar u}}_t} = {\bf{V}}{{\bf{\Sigma }}^{{1 \mathord{\left/
 {\vphantom {1 2}} \right.
 \kern-\nulldelimiterspace} 2}}}{{\bf{r}}_t},t = 1,2,...,T.
\end{equation}
Therefore, the candidate reflection matrix can be expressed as
\begin{equation}
{{\bf{\Theta }}_t} = {\rm{diag}}\left( {\left. {{e^{j\arg \left( {\frac{{{{{\bf{\bar u}}}_t}\left[ m \right]}}{{{{{\bf{\bar u}}}_t}\left[ {M + 1} \right]}}} \right)}}} \right|\forall m \in {\cal M}} \right),
\end{equation}
where ${{\bf{\bar u}}_t}\left[ m \right]$ represents the $m$-th reflecting element of ${{\bf{\bar u}}_t}$. According to the obtained candidate reflection matrix set $\left\{ {\left. {{{\bf{\Theta }}_t}} \right|t = 1,2,...,T} \right\}$, we can obtain the one that maximizes the combined channel gain of all users, i.e.,
\begin{equation}
{t^*} = \arg \mathop {\max }\limits_t \sum\limits_{k = 1}^K {{{\left\| {{\bf{h}}_{r,k}^H{{\bf{\Theta }}_t}{\bf{G}} + {\bf{h}}_{d,k}^H} \right\|}^2}}  .
\end{equation}
It has been verified that SDR technique followed by such randomization scheme can guarantee at least a $\pi/4$-approximation of the optimal objective value of the problem (P2) \cite{8811733}. 


\subsection{BS Beamforming Vector Optimization}
Let ${\bf{h}}_k^H = {\bf{h}}_{r,k}^H{\bf{\Theta G}} + {\bf{h}}_{d,k}^H \in {\mathbb{C}^{1 \times N}}$. Based on the SIC decoding order obtained in the first stage, and given PS ratio IRS phase shift, the problem (P1) can be transformed into the problem (P3), which can be expressed as
\begin{subequations}
	\begin{align}
	\left( {{\textrm{P3}}} \right){\rm{~~~~~}}& \mathop {\min }\limits_{\left\{ {{{\bf{w}}_k}} \right\}} {\rm{~~}}\sum\limits_{k = 1}^K {{{\left\| {{{\bf{w}}_k}} \right\|}^2},} \\
	\rm{s.t.}\qquad &\frac{{{\rho _k}{{\left| {{{\bf{h}}_k^H}{{\bf{w}}_k}} \right|}^2}}}{{{\rho _k}\sum\limits_{s \left( j \right) > s \left( k \right)} {{{\left| {{{\bf{h}}_k^H}{{\bf{w}}_j}} \right|}^2}}  + {\rho _k}\sigma _k^2 + \delta _k^2}} \ge {\gamma _k},\\
	&\frac{{{\rho _k}{{\left| {{\bf{h}}_k^H{{\bf{w}}_k}} \right|}^2}}}{{{\rho _k}\sum\limits_{s\left( j \right) > s\left( k \right)} {{{\left| {{\bf{h}}_k^H{{\bf{w}}_j}} \right|}^2}}  + {\rho _k}\sigma _k^2 + \delta _k^2}} \le \frac{{{\rho _{\bar k}}{{\left| {{\bf{h}}_{\bar k}^H{{\bf{w}}_k}} \right|}^2}}}{{{\rho _{\bar k}}\sum\limits_{s\left( j \right) > s\left( k \right)} {{{\left| {{\bf{h}}_{\bar k}^H{{\bf{w}}_j}} \right|}^2}}  + {\rho _{\bar k}}\sigma _{\bar k}^2 + \delta _{\bar k}^2}},~{\rm{if}}~{\rm{ }}s\left( k \right) < s\left( {\bar k} \right),\\
	&{\eta _k}\left( {1 - {\rho _k}} \right)\left( {\sum\limits_{j = 1}^K {{{\left| {{{\bf{h}}_k^H}{{\bf{w}}_j}} \right|}^2}}  + \sigma _k^2} \right) \ge {e_k}.
	\end{align}
\end{subequations}
Define ${{\bf{W}}_k}{\rm{ = }}{{\bf{w}}_k}{\bf{w}}_k^H \in {\mathbb{C}^{N \times N}},\forall k$, and ${{\bf{W}}_k}$ satisfies ${\rm{Rank}}\left( {{{\bf{W}}_k}} \right) = {\rm{1}},\forall k $. We first apply SDR to relax the rank-one constraint, the problem (P3) can be transformed into the problem (P3.1), which can be expressed as
\begin{subequations}
	\begin{align}
	\left( {{\textrm{P3.1}}} \right){\rm{~~~~}}&\mathop {\min }\limits_{\left\{ {{{\bf{W}}_k}} \right\}} \;\;\sum\limits_{k = 1}^K {{\rm{Tr}}\left( {{{\bf{W}}_k}} \right)} ,  \\
	\rm{s.t.}\qquad &\frac{{{\bf{h}}_k^H{{\bf{W}}_k}{{\bf{h}}_k}}}{{\sum\limits_{s\left( j \right) > s\left( k \right)} {{\bf{h}}_k^H{{\bf{W}}_j}{{\bf{h}}_k}}  + \sigma _k^2 + \frac{{\delta _k^2}}{{{\rho _k}}}}} \ge {\gamma _k},\\
	&\frac{{{\bf{h}}_k^H{{\bf{W}}_k}{{\bf{h}}_k}}}{{\sum\limits_{s\left( j \right) > s\left( k \right)} {{\bf{h}}_k^H{{\bf{W}}_j}{{\bf{h}}_k}}  + \sigma _k^2 + \frac{{\delta _k^2}}{{{\rho _k}}}}} \le \frac{{{\bf{h}}_{\bar k}^H{{\bf{W}}_k}{{\bf{h}}_{\bar k}}}}{{\sum\limits_{s\left( j \right) > s\left( k \right)} {{\bf{h}}_{\bar k}^H{{\bf{W}}_j}{{\bf{h}}_{\bar k}}}  + \sigma _{\bar k}^2 + \frac{{\delta _{\bar k}^2}}{{{\rho _{\bar k}}}}}},~{\rm{if}}~{\rm{ }}s\left( k \right) < s\left( {\bar k} \right),\\
	&{\eta _k}\left( {1 - {\rho _k}} \right)\left( {\sum\limits_{j = 1}^K {{\bf{h}}_k^H{{\bf{W}}_j}{{\bf{h}}_k}}  + \sigma _k^2} \right) \ge {e_k},\\
	&{{\bf{W}}_k} \succeq 0.
	\end{align}
\end{subequations}
Since the constraint (26c) is non-convex, the problem (P3.1) is still a non-convex optimization problem. In this paper, we use SCA to convert the constraint (26c) into
\begin{equation}
	{\hat f_1}\left( {{{\bf{W}}_k}} \right) - \ln \left( {\sum\limits_{s\left( j \right) > s\left( k \right)} {{\bf{h}}_k^H{{\bf{W}}_j}{{\bf{h}}_k}}  + {A_k}} \right) - \ln \left( {{\bf{h}}_{\bar k}^H{{\bf{W}}_k}{{\bf{h}}_{\bar k}}} \right) + {\hat f_2}\left( {{{\bf{W}}_j}} \right) \le 0,~{\rm{if}}~{\rm{ }}s\left( k \right) < s\left( {\bar k} \right),
\end{equation}
where ${{\bf{W}}_k^{\left( r \right)}}$ is the value of ${\bf{W}}_k$ for the $r$-th iteration. See Appendix A for ${\hat f_1}\left( {{{\bf{W}}_k}} \right)$ and ${\hat f_2}\left( {{{\bf{W}}_j}} \right)$. 

Please refer to Appendix A for the detailed proof. $\hfill\blacksquare$

Accordingly, the problem (P3.1) can be transformed into the problem (P3.2), which can be expressed as 
\begin{subequations}
	\begin{align}
		\left( {{\textrm{P3.2}}} \right){\rm{~~~~}}&\mathop {\min }\limits_{\left\{ {{{\bf{W}}_k}} \right\}} \;\;\sum\limits_{k = 1}^K {{\rm{Tr}}\left( {{{\bf{W}}_k}} \right)} ,  \\
		\rm{s.t.}\qquad &\textrm {(26b), (27), (26d), (26e).}
	\end{align}
\end{subequations}
The problem (P3.2) is a standard SDP problem, which can be solved by applying the CVX toolbox \cite{grant2014cvx}. We assume that ${\bf{W}}_k^ * $ is the optimal solution of the problem (P3.2). If ${\bf{W}}_k^ * $ satisfies ${\rm{Rank}}\left( {{\bf{W}}_k^ * } \right) = 1,\forall k$, the beamforming vector ${\bf{w}}_k^ * $ of the problem (P3) can be obtained by eigenvalue decomposition. Next, we have the following proposition.

$\textbf{Proposition~1:}$ The optimal solution ${\bf{W}}_k^ * $ of the problem (P3.2) satisfies ${\rm{Rank}}\left( {{\bf{W}}_k^ * } \right) = 1,\forall k$, i.e., the SDR for the problem (P3) is tight. 

Please refer to Appendix B for the detailed proof.$\hfill\blacksquare$


\subsection{PS Ratio Optimization}
In this sub-problem, according to the SIC decoding order obtained in the first stage, given BS beamforming vector and IRS phase shift, the problem (P1) can be transformed into a feasibility-check problem (P4), which can be given by
\begin{subequations}
	\begin{align}
		\left( {{\textrm{P4}}} \right){\rm{~~~~~}}&{{\rm{find~~}}{\rho _k},} \\
		\rm{s.t.}\qquad &{\frac{{{\rho _k}{{\left| {{\bf{h}}_k^H{{\bf{w}}_k}} \right|}^2}}}{{{\rho _k}\left( {\sum\limits_{s\left( j \right) > s\left( k \right)} {{{\left| {{\bf{h}}_k^H{{\bf{w}}_j}} \right|}^2}}  + \sigma _k^2} \right) + \delta _k^2}} \ge {\gamma _k}},\\
		&{\frac{{{\rho _k}{{\left| {{\bf{h}}_k^H{{\bf{w}}_k}} \right|}^2}}}{{{\rho _k}\left( {\sum\limits_{s\left( j \right) > s\left( k \right)} {{{\left| {{\bf{h}}_k^H{{\bf{w}}_j}} \right|}^2}}  + \sigma _k^2} \right) + \delta _k^2}} \le \frac{{{\rho _{\bar k}}{{\left| {{\bf{h}}_{\bar k}^H{{\bf{w}}_k}} \right|}^2}}}{{{\rho _{\bar k}}\left( {\sum\limits_{s\left( j \right) > s\left( k \right)} {{{\left| {{\bf{h}}_{\bar k}^H{{\bf{w}}_j}} \right|}^2}}  + \sigma _{\bar k}^2} \right) + \delta _{\bar k}^2}}},~{\rm{if}}~{\rm{ }}s\left( k \right) < s\left( {\bar k} \right),\\
		&{{\eta _k}\left( {1 - {\rho _k}} \right)\left( {\sum\limits_{j = 1}^K {{{\left| {{\bf{h}}_k^H{{\bf{w}}_j}} \right|}^2}}  + \sigma _k^2} \right) \ge {e_k}},\\
		&{{\rm{0}} \le {\rho _k} \le 1.}
	\end{align}
\end{subequations}
Due to the existence of the non-convex constraint (29c), the problem (P4) is a non-convex optimization problem. By applying SCA, the constraint (29c) is transformed into
\begin{equation}
	\frac{{{B_{kk}}\delta _{\bar k}^2}}{{{\rho _{\bar k}}}} - {B_{\bar kk}}\delta _k^2\left( {\frac{1}{{\rho _k^{\left( r \right)}}} - \frac{1}{{{{\left( {\rho _k^{\left( r \right)}} \right)}^2}}}\left( {{\rho _k} - \rho _k^{\left( r \right)}} \right)} \right) \le {B_{\bar kk}}{B_{jk}} - {B_{kk}}{B_{j\bar k}},~{\rm{if}}~{\rm{ }}s\left( k \right) < s\left( {\bar k} \right),
\end{equation}
where $\rho _k^{\left( r \right)}$ is the value of $\rho_k$ for the $r$-th iteration. 

Please refer to the Appendix C for detailed proof. $\hfill\blacksquare$

Thus, the problem (P4) can be rewritten as 
\begin{subequations}
	\begin{align}
		\left( {{\textrm{P4.1}}} \right){\rm{~~~~}}&{{\rm{find~~}}{\rho _k},}  \\
		\rm{s.t.}\qquad &\textrm {(29b), (30), (29d), (29e).}
	\end{align}
\end{subequations}
It can be seen that the problem (P4.1) is a standard convex optimization problem, which can be solved by applying the CVX toolbox \cite{grant2014cvx}. 
\subsection{IRS Phase Shift Optimization}
When the SIC decoding order is determined, and beamforming vector and PS ratio are fixed, according to the variable substitution ${\bf{u}}{\rm{ = }}{\left[ {{e^{j{\theta _1}}},...,{e^{j{\theta _M}}}} \right]^H} \in {\mathbb{C}^{M \times 1}}$ in Section III.A, the constraint on ${\theta _m}$ can be transformed into $\left| {{u_m}} \right| = 1,\forall m \in {\cal M}$. Let ${{\bf{p}}_{k,j}} = {\rm{diag}}\left( {{\bf{h}}_{r,j}^H} \right){\bf{G}}{{\bf{w}}_k} \in {\mathbb{C}^{M \times 1}}$, ${{\bf{q}}_{k,j}} = {\bf{h}}_{d,j}^H{{\bf{w}}_k} \in \mathbb{C}$. Then ${\left| {\left( {{\bf{h}}_{r,j}^H{\bf{\Theta G}} + {\bf{h}}_{d,j}^H} \right){{\bf{w}}_k}} \right|^2} = {\left| {{{\bf{u}}^H}{{\bf{p}}_{k,j}} + {{\bf{q}}_{k,j}}} \right|^2}$. Next we introduce auxiliary variables as follows,
\begin{equation}
{{\bf{S}}_{k,j}} = \left[ {\begin{array}{*{20}{c}}
{{{\bf{p}}_{k,j}}{\bf{p}}_{k,j}^H}&{{{\bf{p}}_{k,j}}{\bf{q}}_{k,j}^H}\\
{{{\bf{q}}_{k,j}}{\bf{p}}_{k,j}^H}&0
\end{array}} \right],{\bf{\bar u}}{\rm{ = }}\left[ {\begin{array}{*{20}{c}}
{\bf{u}}\\
1
\end{array}} \right].
\end{equation}
Then ${\left| {{{\bf{u}}^H}{{\bf{p}}_{k,j}} + {{\bf{q}}_{k,j}}} \right|^2} = {{\bf{\bar u}}^H}{{\bf{S}}_{k,j}}{\bf{\bar u}} + {\left| {{{\bf{q}}_{k,j}}} \right|^2}$. Since ${{\bf{\bar u}}^H}{{\bf{S}}_{k,j}}{\bf{\bar u}} = {\rm{Tr}}\left( {{{\bf{S}}_{k,j}}{\bf{\bar u}}{{{\bf{\bar u}}}^H}} \right)$, we define ${\bf{\bar U}}{\rm{ = }}{\bf{\bar u}}{{\bf{\bar u}}^H}$, where ${\bf{\bar U}}$ satisfies ${\bf{\bar U}} \succeq 0$ and ${\rm{Rank}}\left( {\bf{\bar U}} \right) = 1$. Since rank-one constraint is non-convex constraint, we firstly apply the SDR to relax it. Because the objective function of the problem (P1) does not contain the IRS phase shift variable, the problem (P1) is transformed into a feasibility-check problem (P5), which can be expressed as
\begin{subequations}
	\begin{align}
	\left( {{\textrm{P5}}} \right){\rm{~~~~~}}&{\rm{find   }}~~~{\bf{\bar U}}, \\
	\rm{s.t.}\qquad &\frac{{{\rm{Tr}}\left( {{{\bf{S}}_{k,k}}{\bf{\bar U}}} \right) + {{\left| {{{\bf{q}}_{k,k}}} \right|}^2}}}{{\sum\limits_{s\left( j \right) > s\left( k \right)} {\left( {{\rm{Tr}}\left( {{{\bf{S}}_{j,k}}{\bf{\bar U}}} \right) + {{\left| {{{\bf{q}}_{j,k}}} \right|}^2}} \right) + \sigma _k^2 + \frac{{\delta _k^2}}{{{\rho _k}}}} }} \ge {\gamma _k},\\
	&\frac{{{\rm{Tr}}\left( {{{\bf{S}}_{k,k}}{\bf{\bar U}}} \right) + {{\left| {{{\bf{q}}_{k,k}}} \right|}^2}}}{{\sum\limits_{s\left( j \right) > s\left( k \right)} {\left( {{\rm{Tr}}\left( {{{\bf{S}}_{j,k}}{\bf{\bar U}}} \right) + {{\left| {{{\bf{q}}_{j,k}}} \right|}^2}} \right) + \sigma _k^2 + \frac{{\delta _k^2}}{{{\rho _k}}}} }} \le \frac{{{\rm{Tr}}\left( {{{\bf{S}}_{k,\bar k}}{\bf{\bar U}}} \right) + {{\left| {{{\bf{q}}_{k,\bar k}}} \right|}^2}}}{{\sum\limits_{s\left( j \right) > s\left( k \right)} {\left( {{\rm{Tr}}\left( {{{\bf{S}}_{j,\bar k}}{\bf{\bar U}}} \right) + {{\left| {{{\bf{q}}_{j,\bar k}}} \right|}^2}} \right) + \sigma _{\bar k}^2 + \frac{{\delta _{\bar k}^2}}{{{\rho _{\bar k}}}}} }},\\
	&{\rm{Tr}}\left( {{{\bf{R}}_k}{\bf{\bar U}}} \right) + {\left\| {{\bf{h}}_{d,k}^H} \right\|^2} \le {\rm{Tr}}\left( {{{\bf{R}}_{\bar k}}{\bf{\bar U}}} \right) + {\left\| {{\bf{h}}_{d,\bar k}^H} \right\|^{\rm{2}}},\\
	&{{\eta _k}\left( {1 - {\rho _k}} \right)\left( {\sum\limits_{j = 1}^K {{\rm{Tr}}\left( {{{\bf{S}}_{j,k}}{\bf{\bar U}}} \right) + {{\left| {{{\bf{q}}_{j,k}}} \right|}^2} + \sigma _k^2} } \right) \ge {e_k}},\\
	&{{\bf{\bar U}}_{m,m}} = 1,m = 1,2,...,M + 1,\\
	&{\rm{ }}{\bf{\bar U}} \succeq 0.
	\end{align}
\end{subequations}
We apply SCA to transform the non-convex constraint (33c) into
\begin{equation}
	{\hat g_1}\left( {{\bf{\bar U}}} \right) - \ln \left( {\sum\limits_{s\left( j \right) > s\left( k \right)} {\left( {{\rm{Tr}}\left( {{{\bf{S}}_{j,k}}{\bf{\bar U}}} \right) + {{\left| {{{\bf{q}}_{j,k}}} \right|}^2}} \right) + {C_k}} } \right) - \ln \left( {{\rm{Tr}}\left( {{{\bf{S}}_{k,\bar k}}{\bf{\bar U}}} \right) + {{\left| {{{\bf{q}}_{k,\bar k}}} \right|}^2}} \right) + {\hat g_2}\left( {{\bf{\bar U}}} \right) \le 0,
\end{equation}
where ${{{{\bf{\bar U}}}^{\left( r \right)}}}$ is the value of ${{\bf{\bar U}}}$ for the $r$-th iteration. See Appendix D for ${{\hat g}_1}\left( {{\bf{\bar U}}} \right)$ and ${{\hat g}_2}\left( {{\bf{\bar U}}} \right)$. 

Please refer to Appendix D for the detailed proof. $\hfill\blacksquare$

Thus, the problem (P5) can be transformed into the problem (P5.1), which can be expressed as
\begin{subequations}
	\begin{align}
		\left( {{\textrm{P5.1}}} \right){\rm{~~~~}}&{\rm{find   }}~~~{\bf{\bar U}},  \\
		\rm{s.t.}\qquad &\textrm {(33b), (34), (33d), (33e), (33f), (33g).}
	\end{align}
\end{subequations}
The problem (P5.1) is a standard SDP programming problem, which can be solved by CVX toolbox \cite{grant2014cvx}. In general, problem (P5.1) does not produce a rank-one solution, i.e., ${\rm{Rank}}\left( {\bf{U}} \right) \ne 1$. The optimal solution obtained by the problem (P5.1) is only the upper bound of the optimal solution. Therefore, it is necessary to reconstruct the high-rank solution obtained in problem (P5.1) into a rank-one solution, i.e., to reduce the rank of the high-rank solution by using the Gaussian randomization in the subsection III.A. 
%
%

\subsection{Two-stage Overall Optimization Algorithm}
Based on the previous sub-sections, the two-stage overall optimization algorithm is summarized as $\textbf{Algorithm 1}$ (i.e. JDBPR algorithm). Specifically, in the first stage, the SIC decoding order is determined by solving the problem (P2), which mainly optimizes the IRS phase shift matrix to maximize the combined channel gain of all users, and then determines the decoding order according to the combined channel gain of each user. Based on the SIC decoding order obtained in the first stage, when PS ratio and IRS phase shift are fixed in the second stage, the BS beamforming vector is obtained by solving the problem (P3) by applying SDR and SCA, and SDR is proved tight. Next, the last two sub-problems of PS ratio and IRS phase shift can be obtained by solving problem (P4) and (P5) by using SDR, SCA and Gaussian randomization when BS transmit beamforming vector is given. The three sub-problems of the second stage are alternately optimized to achieve convergence\footnote{For the two-stage overall optimization algorithm (i.e., JDBPR algorithm) proposed in this paper, the Gaussian randomization used in the first stage can obtain an approximate value of at least $\pi/4$ for the optimal objective value of this stage. Meanwhile, for the three sub-problems of the second stage, since each of the three sub-problems has a complex SIC decoding condition constraint, we apply SCA to solve these sub-problems. Therefore, the JDBPR algorithm can obtain a sub-optimal solution of the problem (P1).}

\begin{algorithm}[H]
	\caption{Two-Stage $\textbf{J}$oint SIC $\textbf{D}$ecoding Order, BS Transmit $\textbf{B}$eamforming Vector, $\textbf{P}$S Ratio and I$\textbf{R}$S Phase Shift Optimization (JDBPR) Algorithm} 
	\begin{algorithmic}[1]
		\State {\bf{Stage 1:}} SIC decoding order determination.
		\State Obtain the SIC decoding order $\left\{ {s\left( k \right)} \right\}$ by solving the problem (P2.1).
		\State {\bf{Stage 2:}} Joint BS transmit beamforming vector, PS ratio and IRS phase shift optimization.
		\State Initialize ${\left\{ {{{\bf{w}}_k}} \right\}^{\left( 0 \right)}}$, ${\left\{ {{\rho _k}} \right\}^{\left( 0 \right)}}$ and ${\left\{ {{\theta _m}} \right\}^{\left( 0 \right)}}$. Let $r=0$, $\varepsilon  = {10^{ - 3}}$.
	    \Repeat
	    \State Solve the problem (P3.2) for given ${\left\{ {{\rho _k}} \right\}^{\left( r \right)}}$ and ${\left\{ {{\theta _m}} \right\}^{\left( r \right)}}$, and obtain BS transmit beamforming vector ${\left\{ {{{\bf{w}}_k}} \right\}^{\left( {r + 1} \right)}}$.
	    \State Solve the problem (P4.1) for given ${\left\{ {{{\bf{w}}_k}} \right\}^{\left( r + 1 \right)}}$ and ${\left\{ {{\theta _m}} \right\}^{\left( r \right)}}$, and obtain PS ratio ${\left\{ {{\rho _k}} \right\}^{\left( {r + 1} \right)}}$.
	    \State Solve the problem (P5.1) for given ${\left\{ {{{\bf{w}}_k}} \right\}^{\left( r + 1 \right)}}$ and ${\left\{ {{\rho _k}} \right\}^{\left( {r + 1} \right)}}$, and obtain IRS phase shift ${\left\{ {{\theta _m}} \right\}^{\left( r + 1 \right)}}$.
	    \State Update $r=r+1$.
	    \Until  The fractional decrease of the objective value is below a threshold $\varepsilon$.
		\State \Return SIC decoding order, BS transmit beamforming vector, PS ratio and IRS phase shift.
	\end{algorithmic}
\end{algorithm}

\subsection{Computational Complexity and Convergence Analysis of the Proposed JDBPR Algorithm}
\subsubsection{Computational complexity analysis}
In each iteration, since both problem (P2.1) and (5.1) optimize the IRS phase shift, they both solve a relaxed SDP problem by interior point method, so the computational complexity of problem (P2.1) and (5.1) in solving the SDP problem can be represented by ${\cal O}\left( {{{\left( {M + 1} \right)}^{3.5}}} \right)$. The computational complexity of using the interior point method to solve the problem (P3.2) is ${\cal O}\left( {K{N^{3.5}}} \right)$ \cite{ben2001lectures}, and the problem (P4.1) is solved with the complexity of ${\cal O}\left( K \right)$. We assume that the number of iterations required for the algorithm to reach convergence is $r$, the computational complexity of the proposed JDBPR algorithm can be expressed as ${\cal O}\left( {r\left( {K{N^{3.5}} + {{(M + 1)}^{3.5}} + K} \right)} \right)$.
\subsubsection{Convergence analysis}
The convergence of the proposed two-stage JDBPR algorithm mainly lies in the second stage because the first stage only determines the decoding order. Therefore, the convergence performance of the second-stage problem needs to be proved as follows.

We define ${\left\{ {{{\bf{w}}_k}} \right\}^{\left( {r} \right)}}$, ${\left\{ {{\rho _k}} \right\}^{\left( r \right)}}$ and ${\left\{ {{\theta _m}} \right\}^{\left( r \right)}}$ as the $r$-th iteration solution of the problem (P3.2), (P4.1) and (P5.1). The objective function is denoted by ${\cal P}\left( {{{\left\{ {{{\bf{w}}_k}} \right\}}^{\left( r \right)}},{{\left\{ {{\rho _k}} \right\}}^{\left( r \right)}},{{\left\{ {{\theta _m}} \right\}}^{\left( r \right)}}} \right)$. In the step 6 of $\textbf{Algorithm 1}$, since BS transmit beamforming vector can be obtained for given ${\left\{ {{\rho _k}} \right\}^{\left( r \right)}}$ and ${\left\{ {{\theta _m}} \right\}^{\left( r \right)}}$. Hence, we have 
\begin{equation}
{\cal P}\left( {{{\left\{ {{{\bf{w}}_k}} \right\}}^{\left( r \right)}},{{\left\{ {{\rho _k}} \right\}}^{\left( r \right)}},{{\left\{ {{\theta _m}} \right\}}^{\left( r \right)}}} \right) \ge {\cal P}\left( {{{\left\{ {{{\bf{w}}_k}} \right\}}^{\left( {r + 1} \right)}},{{\left\{ {{\rho _k}} \right\}}^{\left( r \right)}},{{\left\{ {{\theta _m}} \right\}}^{\left( r \right)}}} \right).
\end{equation}
In the step 7 of $\textbf{Algorithm 1}$, we can obtain PS ratio when ${\left\{ {{{\bf{w}}_k}} \right\}^{\left( r + 1 \right)}}$ and ${\left\{ {{\theta _m}} \right\}^{\left( r \right)}}$ are given. Herein, we also have 
\begin{equation}
{\cal P}\left( {{{\left\{ {{{\bf{w}}_k}} \right\}}^{\left( {r + 1} \right)}},{{\left\{ {{\rho _k}} \right\}}^{\left( r \right)}},{{\left\{ {{\theta _m}} \right\}}^{\left( r \right)}}} \right) \ge {\cal P}\left( {{{\left\{ {{{\bf{w}}_k}} \right\}}^{\left( {r + 1} \right)}},{{\left\{ {{\rho _k}} \right\}}^{\left( {r + 1} \right)}},{{\left\{ {{\theta _m}} \right\}}^{\left( r \right)}}} \right) .
\end{equation}
Similarly, in the step 8 of $\textbf{Algorithm 1}$, we can obtain IRS phase shift when ${\left\{ {{{\bf{w}}_k}} \right\}^{\left( r + 1 \right)}}$ and ${\left\{ {{\rho _k}} \right\}^{\left( {r + 1} \right)}}$ are given. Accordingly,
\begin{equation}
	 {\cal P}\left( {{{\left\{ {{{\bf{w}}_k}} \right\}}^{\left( {r + 1} \right)}},{{\left\{ {{\rho _k}} \right\}}^{\left( {r + 1} \right)}},{{\left\{ {{\theta _m}} \right\}}^{\left( r \right)}}} \right) \ge {\cal P}\left( {{{\left\{ {{{\bf{w}}_k}} \right\}}^{\left( {r + 1} \right)}},{{\left\{ {{\rho _k}} \right\}}^{\left( {r + 1} \right)}},{{\left\{ {{\theta _m}} \right\}}^{\left( {r + 1} \right)}}} \right).
\end{equation}
Based on the above, we have
\begin{equation}
    {\cal P}\left( {{{\left\{ {{{\bf{w}}_k}} \right\}}^{\left( {r } \right)}},{{\left\{ {{\rho _k}} \right\}}^{\left( {r } \right)}},{{\left\{ {{\theta _m}} \right\}}^{\left( r \right)}}} \right) \ge {\cal P}\left( {{{\left\{ {{{\bf{w}}_k}} \right\}}^{\left( {r + 1} \right)}},{{\left\{ {{\rho _k}} \right\}}^{\left( {r + 1} \right)}},{{\left\{ {{\theta _m}} \right\}}^{\left( {r + 1} \right)}}} \right) ,
\end{equation}
which shows that the value of the objective function is non-increasing after each iteration in the second stage of $\textbf{Algorithm 1}$. Since the objective function values of problems (P3.2), (P4.1) and (P5.1) have a finite lower bound, the convergence of $\textbf{Algorithm 1}$ can be guaranteed.

\section{Numerical results}
In this section, we provide simulation results to demonstrate the effectiveness of the proposed JDBPR algorithm in IRS-assisted SWIPT NOMA networks. In this paper, we consider a three-dimensional (3D) coordinate system, where the BS and the IRS placed on the building are located at (0m, 0m, 15m) and (50m, 50m, 15m), respectively, and $K=4$ ground users are randomly and uniformly distributed in a circle whose origin is (0m, 0m, 0m) and a radius of 200m. We consider that the BS is equipped with $N=4$ antennas, and the IRS is equipped with 30 reflecting elements. We assume that the parameters of all users are the same, i.e., $\sigma _k^2=\sigma ^2$, $\delta_k^2=\delta^2$, $\eta_k=\eta$, $\gamma_k=\gamma$ and $e_k=e$. Herein, we set $\sigma ^2=-70$dBm, $\delta^2=-50$dBm and $\eta=0.7$ in our numerical simulations. The path loss exponents are set as ${\alpha}_1=3$, ${\alpha}_2=2.2$ and ${\alpha}_3=2.5$. We set the path loss $C_0$ to $-30$dB when the reference distance is 1m, and we set the Rician factor to 3dB. The number of candidate random vectors used for the Gaussian randomization is set to 1000. The convergence threshold of the JDBPR algorithm is set as $10^{-3}$. The simulation parameter table is shown in Table I.


\newcommand{\tabincell}[2]{\begin{tabular}{@{}#1@{}}#2\end{tabular}}
\begin{table}
	\begin{center}
		\renewcommand{\arraystretch}{1.1}
		\caption{Simulation parameters}
		\label{T1}
		{\begin{tabular}{|c|c|}
			\hline
			\textbf{Parameters}&\textbf{Value}\\
			\hline
			{Number of ground users}&4\\
			\hline
			{Number of antennas}&4\\
			\hline
			{Number of reflecting elements}&30\\
			\hline
			{The radius of circular area}&200m\\
			\hline
			{The coordinates of BS}&(0m, 0m, 15m)\\
			\hline
			{The coordinates of IRS}&(50m, 50m, 15m)\\
			\hline
			{The antenna noise power}&-70dBm\\
			\hline
			{The additional noise power}&-50dBm\\
			\hline
			\tabincell{c}{The path loss with reference\\ distance of 1m}&-30dB\\
			\hline
			{The energy conversion efficiency}&0.7\\
			\hline
			{The path loss exponents}&\tabincell{c}{${\alpha}_1=3$, ${\alpha}_2=2.2$, ${\alpha}_3=2.5$}\\
			\hline
			{The Rician factor}&-3dB\\
			\hline
  				{The number of candidate random vectors}&1000\\
			\hline
			{Convergence threshold}&${10^{ - 3}}$\\
			\hline
		\end{tabular}}
	\end{center}
\end{table}

We first evaluate the convergence performance of the proposed JDBPR algorithm. Fig. 3 shows the BS transmit power varies with the number of iterations under different IRS reflecting elements. It can be seen that as the number of iterations increases, the BS transmit power gradually decreases. The algorithm can reach convergence in the 8-th iteration, which verifies that the proposed algorithm converges fast. Specifically, we compare the performance of the JDBPR algorithm when the number of IRS reflecting elements are 30, 60, and 80 respectively. It can be seen that the greater the number of IRS reflecting elements, the lower the BS transmit power, which also verifies the effectiveness of the IRS assisted SWIPT NOMA networks.

\begin{figure}
	\centerline{\includegraphics[width=7cm]{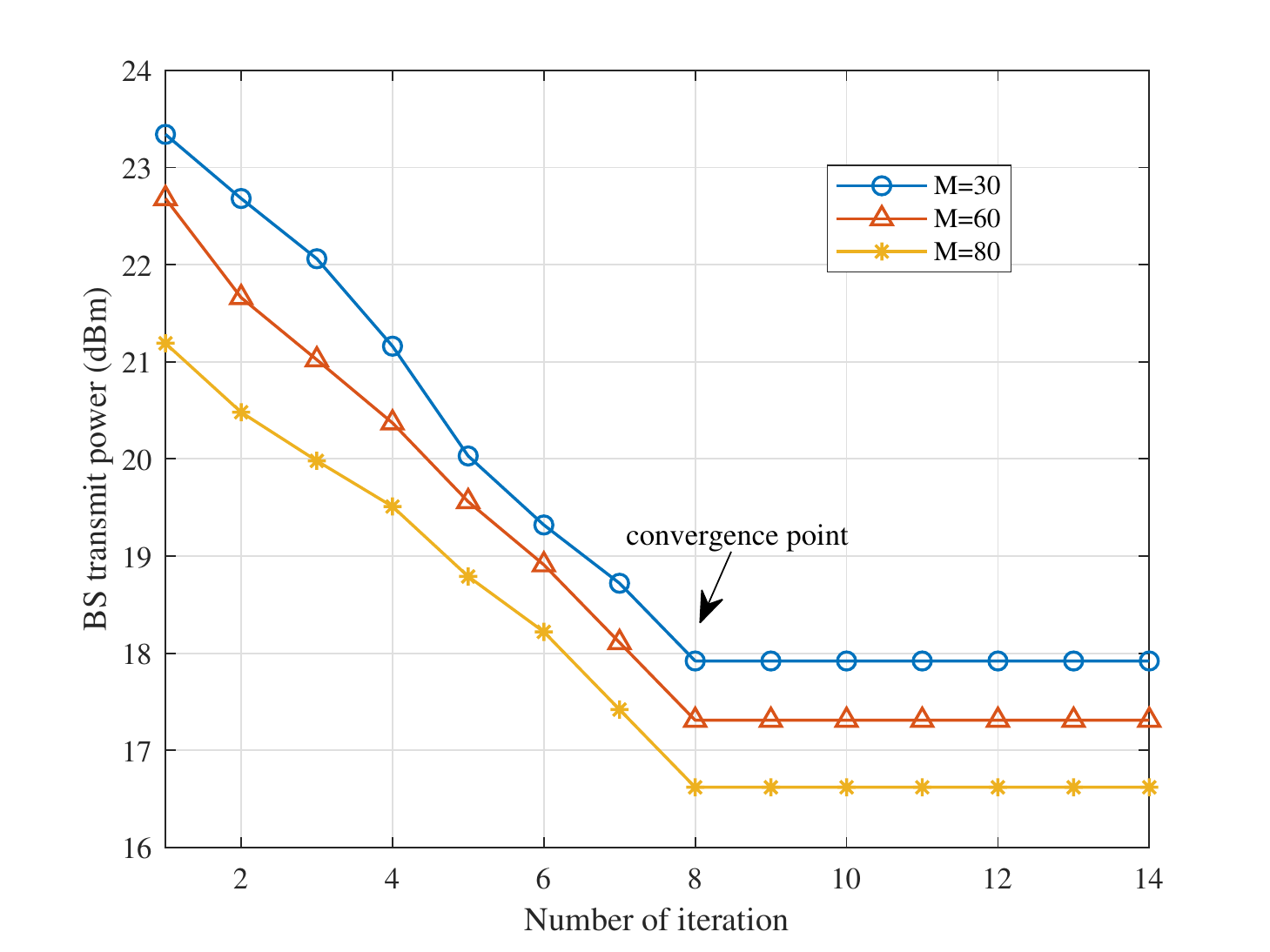}}
	\caption{The convergence of the proposed JDBPR algorithm.}
	\label{Fig3}
\end{figure}

Next, we compare the performance of the proposed JDBPR algorithm with other baseline algorithms. (1) EX-JBPR-opt algorithm: In the first stage, SIC decoding order adopts the exhaustive search method. The algorithm used in the second stage is the same as the second stage of the proposed JDBPR algorithm. (2) TS-JDBPR-opt algorithm: The proposed two-stage JDBPR algorithm. (3) TS-JDBPR-com algorithm: The algorithm is the same as the two stage optimization of the proposed JDBPR algorithm, except that the three sub-problems in the second stage of the former are not optimized alternately. (4) TS-JDBPR-ZF algorithm: The only difference from the TS-JDBPR-opt algorithm is that BS beamforming vector design uses the sub-optimal ZF beamforming scheme, which can eliminate user interference \cite{6805330}. (5) TS-JDBP-ran algorithm: The only difference from the TS-JDBPR-opt algorithm is that the design of IRS phase shift adopts a random scheme. 

Herein, we investigate the behavior of the BS transmit power with the QoS requirements of all users under different energy harvested threshold $e$. Fig. 4 and Fig. 5 respectively show how the BS transmit power varies with the users' QoS requirements under different energy harvested threshold (e.g. $e=0$dBm and $e=-10$dBm). From Fig. 4 and Fig. 5, we can see that the BS transmit power under different algorithms increases with the increase of users' QoS threshold. This is because the larger the users' QoS threshold, the higher the requirements for the BS transmit power. In addition, it can be seen that the performance of the proposed TS-JDBPR-opt algorithm is similar to the EX-JBPR-opt algorithm, but the complexity of exhaustive search is much higher than the proposed algorithm. Meanwhile, our proposed TS-JDBPR-opt algorithm has better performance than the other three baseline algorithms. In specific analysis, the better performance of the TS-JDBPR-opt algorithm compared to the TS-JDBPR-com algorithm is mainly because the former is considered from the perspective of global optimization, while the latter only performs local optimization. Compared with the sub-optimal beamforming scheme of TS-JDPR-ZF algorithm, our proposed beamforming scheme at the BS has better performance. The TS-JDBP-ran algorithm designs the IRS phase shift in a random manner, thus the solution to our proposed algorithm has a better performance.

\begin{figure}[htbp]
	\begin{minipage}[t]{0.5\linewidth}
		\centering
		\includegraphics[width=7cm]{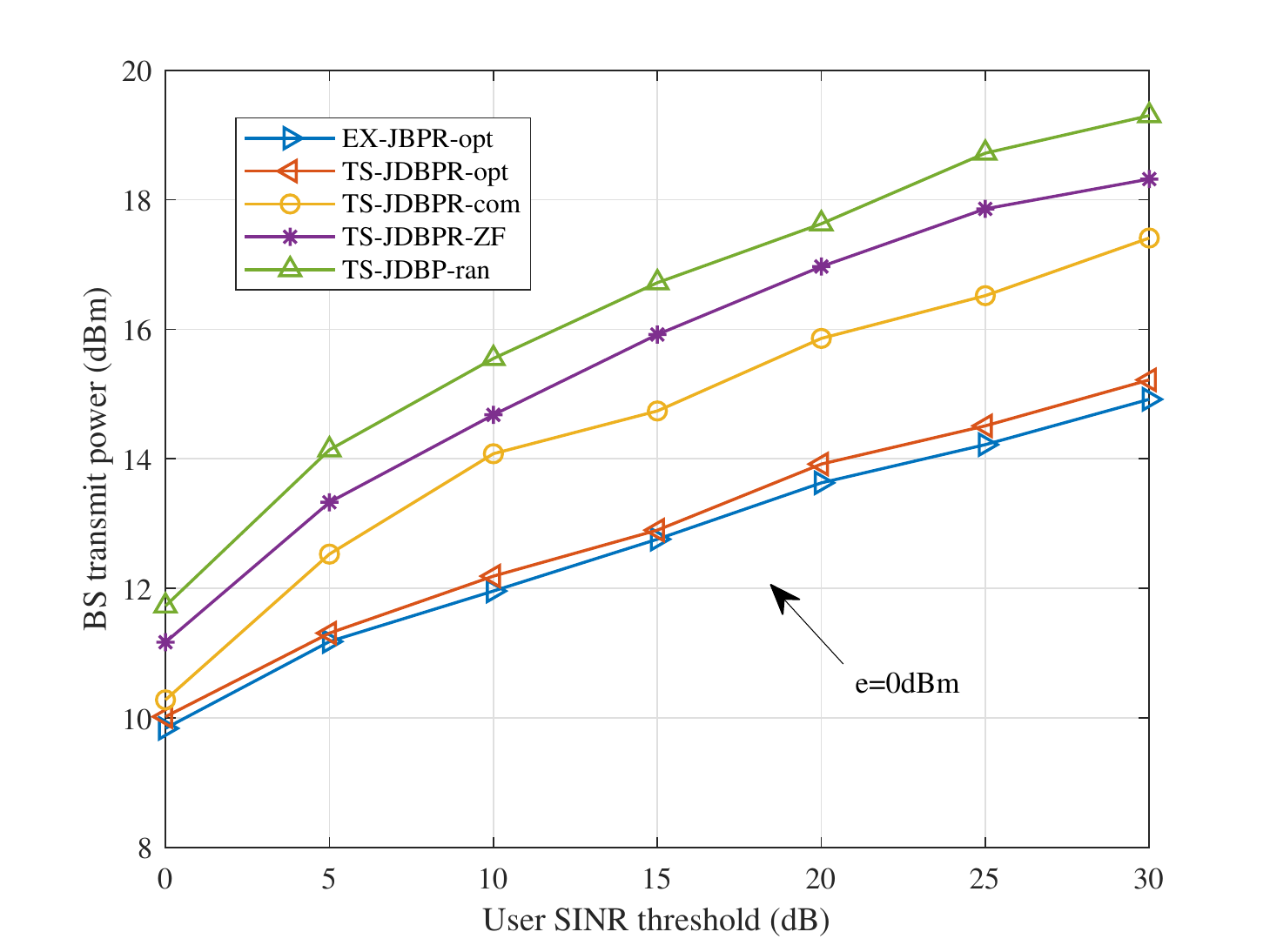}
		\caption{The BS transmit power versus QoS threshold \protect\\when $e=0$dBm.}
	\end{minipage}%
	\begin{minipage}[t]{0.5\linewidth}
		\centering
		\includegraphics[width=7cm]{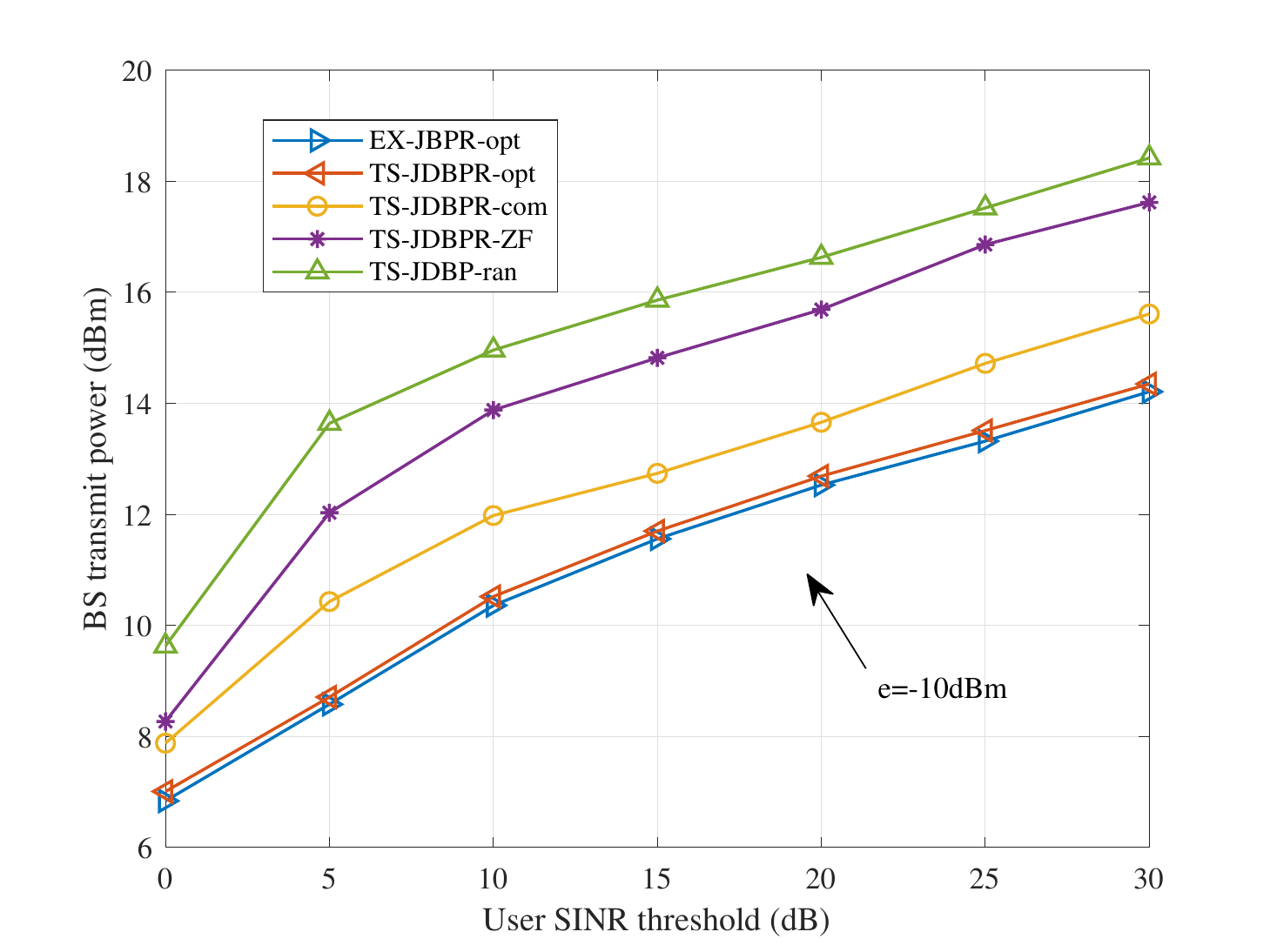}
		\caption{The BS transmit power versus QoS threshold \protect\\when $e=-10$dBm.}
	\end{minipage}
\end{figure}

Fig. 6 and Fig. 7 elaborate how the BS transmit power varies with the user’s energy harvested threshold under different users' QoS threshold (e.g. $\gamma$ = 10dB and $\gamma$ = 0dB). In general, as user energy harvested threshold increases, the BS transmit power continues to increase. This is mainly because the increase in the energy harvested threshold required by the user will require the BS to give a stronger transmission signal. When we consider that the users' QoS threshold is 10dB, it can be seen that the proposed TS-JDBPR-opt algorithm requires similar transmit power to the EX-JBPR-opt algorithm, but its complexity is much lower than the latter. This is mainly because the latter adopts the exhaustive search method in the decoding order scheme, which is more complicated. In addition, the performance of the TS-JDBPR-opt algorithm is also better than the other three baseline algorithms. The reason why the TS-JDBPR-opt algorithm performs better than the TS-JDBPR-com algorithm is that the latter does not achieve the convergence of the entire problem. Compared with the TS-JDBP-ran algorithm, the TS-JDBPR-opt algorithm can greatly reduce the BS transmit power, because that the TS-JDBP-ran algorithm does not optimize the IRS phase shift, and the random phase may even make the system performance worse. In addition, when the users' QoS threshold is 0dB, the change of the BS transmit power with the user's energy harvested threshold is similar to the former. Meanwhile, comparing Fig. 6 and Fig. 7, we can also see that when the user's energy harvested threshold is the same, the higher the users' QoS threshold, the greater the transmit power required by the BS.

\begin{figure}[htbp]
	\begin{minipage}[t]{0.5\linewidth}
		\centering
		\includegraphics[width=7cm]{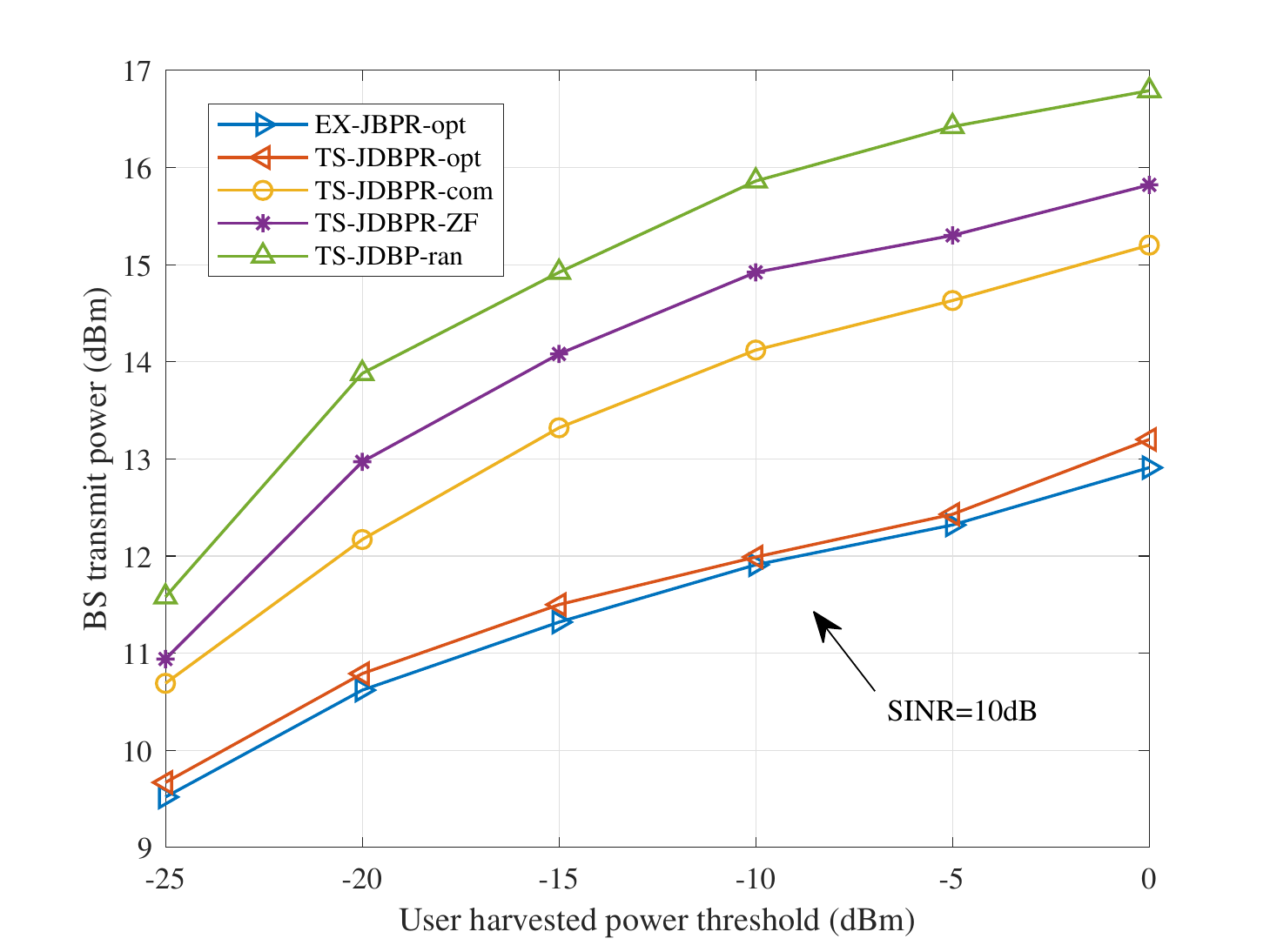}
		\caption{The BS transmit power versus harvested power \protect\\when $\gamma_{th}$ = 10dB.}
	\end{minipage}%
	\begin{minipage}[t]{0.5\linewidth}
		\centering
		\includegraphics[width=7cm]{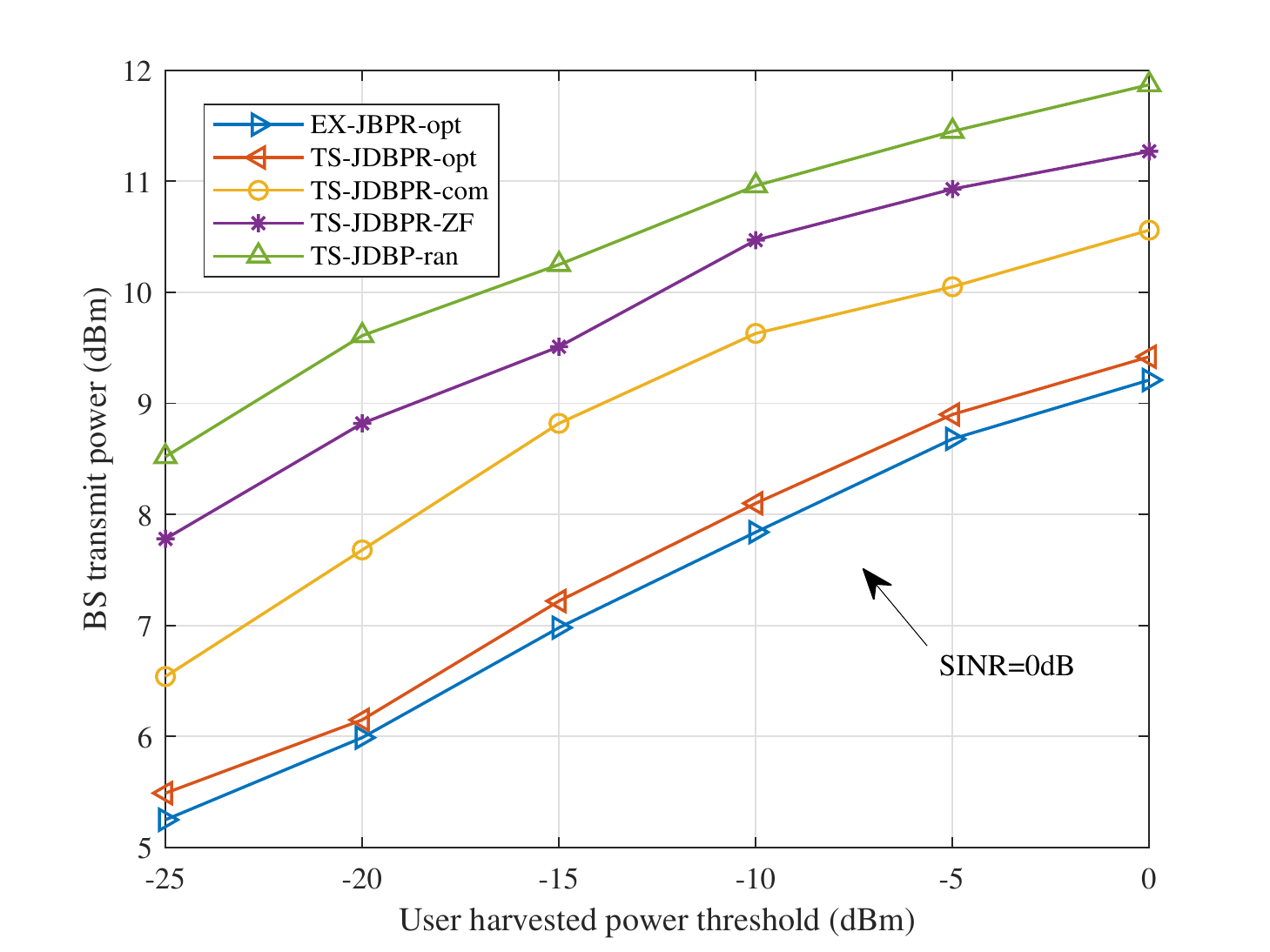}
		\caption{The BS transmit power versus harvested power \protect\\when $\gamma_{th}$ = 0dB.}
	\end{minipage}
\end{figure}

Fig. 8 illustrates the variation of BS transmit power with the number of IRS reflecting elements under different algorithms. In general, it can be seen that as the number of IRS reflecting elements increases, the BS transmit power continues to decrease. This is because that the channel can be adjusted through the IRS, so that the system performance is enhanced, that is, the BS transmit power is gradually reduced. From Fig. 8, it can be seen that the performance of SWIPT NOMA networks with IRS assistance is better than the networks without IRS assistance. Furthermore, when the reflecting elements of IRS increase, the gap of this performance becomes larger, which also verifies that IRS has a very important auxiliary role in SWIPT NOMA networks. At the same time, our proposed TS-JDBPR-opt algorithm achieves a significant performance improvement compared to the TS-JDBPR-com algorithm and the TS-JDBP-ran algorithm. The main reason is that the TS-JDBPR-com algorithm does not achieve global convergence, and the TS-JDBP-ran algorithm does not optimize the IRS phase shift. Fig. 8 can also demonstrate that we can reduce the transmit power on the BS by increasing the number of IRS reflecting elements, and the cost of this scheme is very low. 

\begin{figure}[htbp]
	\begin{minipage}[t]{0.5\linewidth}
		\centering
		\includegraphics[width=7cm]{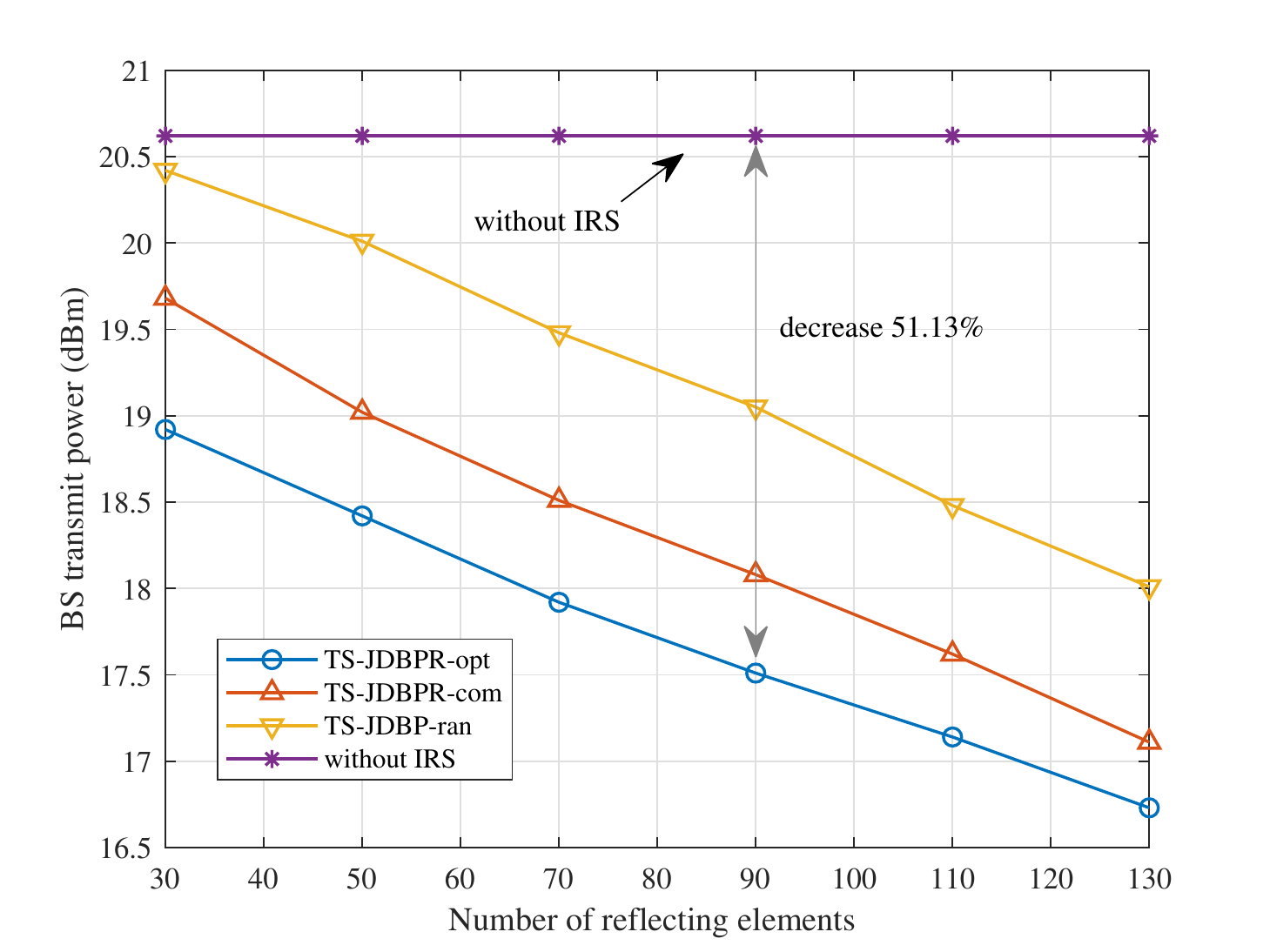}
		\caption{The BS transmit power versus number of \protect \\reflecting elements $M$.}
	\end{minipage}%
	\begin{minipage}[t]{0.5\linewidth}
		\centering
		\includegraphics[width=7cm]{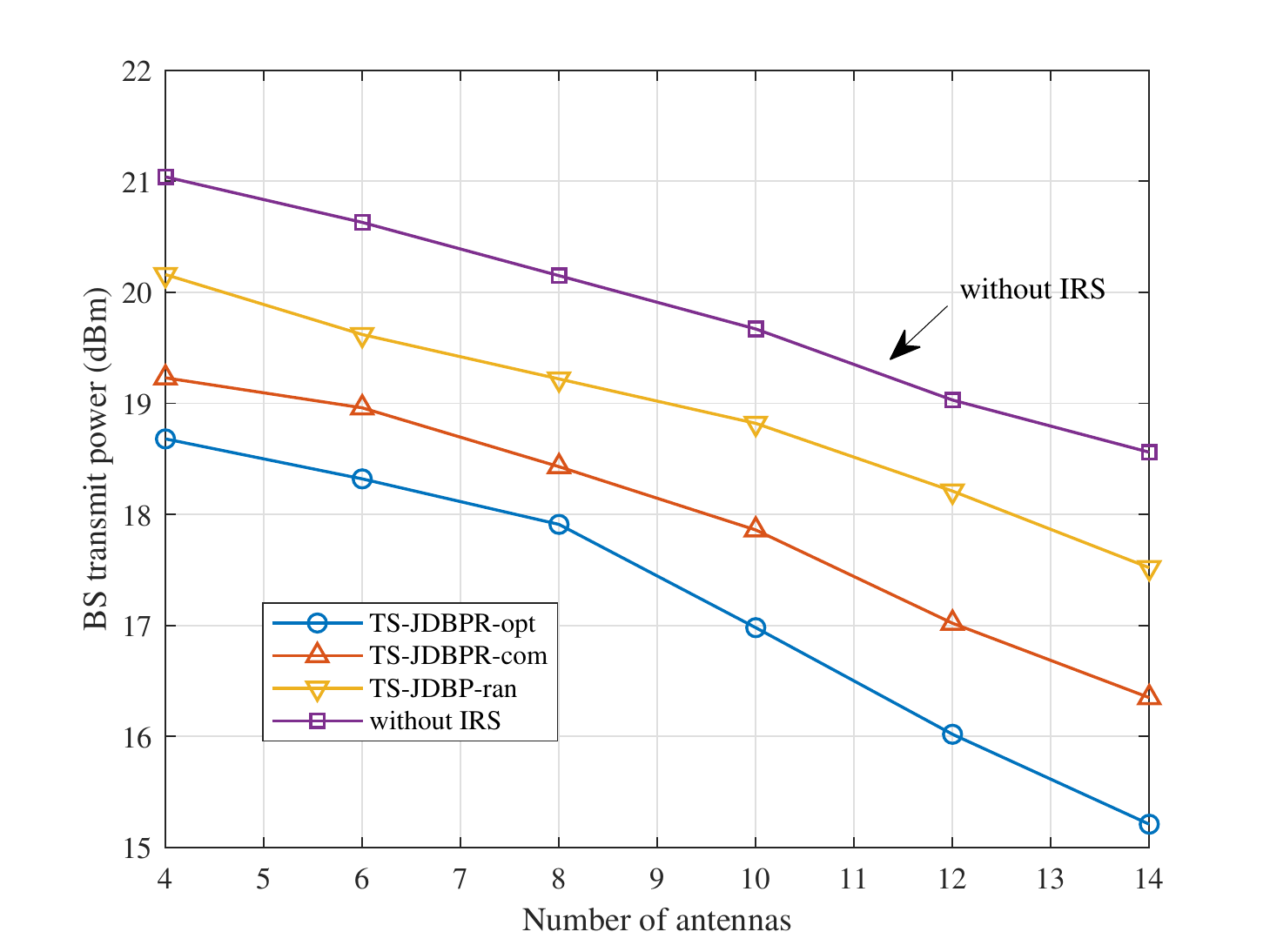}
		\caption{The BS transmit power versus number of \protect \\BS antennas $N$.}
	\end{minipage}
\end{figure}

Finally, Fig. 9 explains how the BS transmit power versus the number of BS antennas under different algorithms. With the increase of the number of antennas, the BS transmit power continues to decrease, which also shows that we can improve the performance of the system by increasing the number of BS antennas. This also motivates us to adopt massive MIMO systems to further enhance the IRS-assisted SWIPT NOMA networks. In addition, when the number of antennas is the same, the proposed TS-JDBPR-opt algorithm still yields a significant performance improvement compared to the TS-JDBPR-com algorithm and the TS-JDBP-ran algorithm.
\section{Conclusion}
This paper investigates the transmit power minimization problem for the IRS-assisted SWIPT NOMA networks. Specifically, under the users' QoS and energy harvested constraints, SIC decoding order, BS transmit beamforming vector, PS ratio, and IRS phase shift are jointly optimized. A two-stage optimization algorithm is proposed to solve this challenging problem. In the first stage, SIC decoding order determination algorithm based on the combined channel gain has been proposed. Further, we divide the second stage into three sub-problems. The BS transmit beamforming vector, PS ratio and IRS phase shift are alternately optimized until convergence is achieved by applying SDR, SCA and  Gaussian randomization. In addition, the computational complexity and convergence of the proposed JDBPR algorithm are analyzed and proved. Numerical results show that our proposed algorithm can significantly reduce the BS transmit power compared to other baseline algorithms, and the auxiliary role of IRS is extremely important, which can greatly relieve the pressure on the BS with low cost in practice.


%

\appendices
\section{Proof of Eq. (27)}
Let ${A_k} = \sigma _k^2 + \frac{{\delta _k^2}}{{{\rho _k}}}$ and ${A_{\bar k}} = \sigma _{\bar k}^2 + \frac{{\delta _{\bar k}^2}}{{{\rho _{\bar k}}}}$. Take the logarithm of both sides of the constraint (26c) as follows
\begin{equation}
	\ln \left( {\frac{{{\bf{h}}_k^H{{\bf{W}}_k}{{\bf{h}}_k}}}{{\sum\limits_{s\left( j \right) > s\left( k \right)} {{\bf{h}}_k^H{{\bf{W}}_j}{{\bf{h}}_k}}  + {A_k}}}} \right) \le \ln \left( {\frac{{{\bf{h}}_{\bar k}^H{{\bf{W}}_k}{{\bf{h}}_{\bar k}}}}{{\sum\limits_{s\left( j \right) > s\left( k \right)} {{\bf{h}}_{\bar k}^H{{\bf{W}}_j}{{\bf{h}}_{\bar k}}}  + {A_{\bar k}}}}} \right),~{\rm{if}}~{\rm{ }}s\left( k \right) < s\left( {\bar k} \right),
\end{equation}
which can also be expressed as
\begin{equation}
	\begin{aligned}
		&\ln \left( {{\bf{h}}_k^H{{\bf{W}}_k}{{\bf{h}}_k}} \right) - \ln \left( {\sum\limits_{s\left( j \right) > s\left( k \right)} {{\bf{h}}_k^H{{\bf{W}}_j}{{\bf{h}}_k}}  + {A_k}} \right)- \ln \left( {{\bf{h}}_{\bar k}^H{{\bf{W}}_k}{{\bf{h}}_{\bar k}}} \right) \\
		& + \ln \left( {\sum\limits_{s\left( j \right) > s\left( k \right)} {{\bf{h}}_{\bar k}^H{{\bf{W}}_j}{{\bf{h}}_{\bar k}}}  + {A_{\bar k}}} \right) \le 0,~{\rm{if}}~{\rm{ }}s\left( k \right) < s\left( {\bar k} \right).
	\end{aligned}
\end{equation}
Since the first term and the fourth term on the left-hand-side (LHS) are both concave, we apply SCA to obtain their upper bound respectively, as follows,
\begin{equation}
	\ln \left( {{\bf{h}}_k^H{{\bf{W}}_k}{{\bf{h}}_k}} \right) \le \ln \left( {{\bf{h}}_k^H{\bf{W}}_k^{\left( r \right)}{{\bf{h}}_k}} \right) + {\rm{Tr}}\left( {{{\left( {\frac{1}{{{\bf{h}}_k^H{\bf{W}}_k^{\left( r \right)}{{\bf{h}}_k}}}{{\bf{h}}_k}{\bf{h}}_k^H} \right)}^H}\left( {{{\bf{W}}_k} - {\bf{W}}_k^{\left( r \right)}} \right)} \right)\buildrel \Delta \over ={\hat f_1}\left( {{{\bf{W}}_k}} \right),
\end{equation}
and
\begin{equation}
	\begin{aligned}
		&\ln \left( {\sum\limits_{s\left( j \right) > s\left( k \right)} {{\bf{h}}_{\bar k}^H{{\bf{W}}_j}{{\bf{h}}_{\bar k}}}  + {A_{\bar k}}} \right) \le \ln \left( {\sum\limits_{s\left( j \right) > s\left( k \right)} {{\bf{h}}_{\bar k}^H{\bf{W}}_j^{\left( r \right)}{{\bf{h}}_{\bar k}}}  + {A_{\bar k}}} \right)  \\
		& + \sum\limits_{s\left( j \right) > s\left( k \right)} {{\rm{Tr}}\left( {{{\left( {\frac{1}{{{\bf{h}}_{\bar k}^H{\bf{W}}_j^{\left( r \right)}{{\bf{h}}_{\bar k}}}}{{\bf{h}}_{\bar k}}{\bf{h}}_{\bar k}^H} \right)}^H}\left( {{{\bf{W}}_j} - {\bf{W}}_j^{\left( r \right)}} \right)} \right)}\buildrel \Delta \over ={\hat f_2}\left( {{{\bf{W}}_j}} \right).
	\end{aligned}
\end{equation}
Therefore, constraint (26c) can be transformed into as follows Eq. (27). The proof is completed.

\section{Proof of the Proposition 1}
First, we introduce auxiliary variables and rewrite the problem (P3.2) as follows 
\begin{subequations}
	\begin{align}
		&{\mathop {\min }\limits_{\left\{ {{{\bf{W}}_k}} \right\}} \;\;\sum\limits_{k = 1}^K {{\rm{Tr}}\left( {{{\bf{W}}_k}} \right)} ,} \\
		\rm{s.t.}\qquad &{{\bf{h}}_k^H{{\bf{W}}_k}{{\bf{h}}_k} - {\gamma _k}\sum\limits_{s\left( j \right) > s\left( k \right)} {{\bf{h}}_k^H{{\bf{W}}_j}{{\bf{h}}_k}}  \ge {\gamma _k}\left( {\sigma _k^2 + \frac{{\delta _k^2}}{{{\rho _k}}}} \right)},\\
		&{{{\hat f}_1}\left( {{{\bf{W}}_k}} \right) - \ln \left( {\sum\limits_{s\left( j \right) > s\left( k \right)} {{\bf{h}}_k^H{{\bf{W}}_j}{{\bf{h}}_k}}  + {A_k}} \right) - \ln \left( {{\bf{h}}_{\bar k}^H{{\bf{W}}_k}{{\bf{h}}_{\bar k}}} \right) + {{\hat f}_2}\left( {{{\bf{W}}_j}} \right) \le 0,{\rm{if}}\;s\left( k \right) < s\left( {\bar k} \right),}\\
		&{\sum\limits_{j = 1}^K {{\bf{h}}_k^H{{\bf{W}}_j}{{\bf{h}}_k}}  + \sigma _k^2 \ge \frac{{{e_k}}}{{{\eta _k}\left( {1 - {\rho _k}} \right)}},}\\
		&{\varphi _k} \le \sum\limits_{s\left( j \right) > s\left( k \right)} {{\bf{h}}_k^H{{\bf{W}}_j}{{\bf{h}}_k}}  + {A_k},\\
		&{\phi _{\bar k}} \le {\bf{h}}_{\bar k}^H{{\bf{W}}_k}{{\bf{h}}_{\bar k}},\\
		&{{\bf{W}}_k} \succeq 0.
	\end{align}
\end{subequations}
Since this problem is convex with respect to (w.r.t) ${\bf{W}}_k$, the Slater's condition holds \cite{boyd2004convex}. Therefore, the duality gap is zero. 
By solving its dual problem, we can obtain its optimal solution. Let ${{\bf{H}}_k} = {{\bf{h}}_k}{\bf{h}}_k^H,\forall k$, the Lagrangian function corresponding to this problem can be given by
\begin{equation}
	\begin{aligned}
		&{\cal L} = \sum\limits_{k = 1}^K {{\rm{Tr}}\left( {{{\bf{W}}_k}} \right)}  - \sum\limits_{k = 1}^K {{\rm{Tr}}\left( {{{\bf{W}}_k}{{\bf{Y}}_k}} \right)}  - \sum\limits_{k = 1}^K {{\lambda _k}} {\rm{Tr}}\left( {{{\bf{W}}_k}{{\bf{H}}_k}} \right) + \\
		&\sum\limits_{s\left( {\bar k} \right) > s\left( k \right)} {{\mu _{k\bar k}}} {\rm{Tr}}\left( {{{\left( {\frac{1}{{{\bf{h}}_k^H{\bf{W}}_k^{\left( r \right)}{{\bf{h}}_k}}}{{\bf{H}}_k}} \right)}^H}\left( {{{\bf{W}}_k} - {\bf{W}}_k^{\left( r \right)}} \right)} \right)- \\ 
		&  \sum\limits_{\bar k = 1}^{K - 1} {{\upsilon _{\bar k}}{\rm{Tr}}\left( {{{\bf{W}}_k}{{\bf{H}}_{\bar k}}} \right)}  -\sum\limits_{k = 1}^K {{\varpi _k}\sum\limits_{j = 1}^K {{\rm{Tr}}\left( {{{\bf{W}}_j}{{\bf{H}}_k}} \right)} }  + \Upsilon ,
	\end{aligned}
\end{equation}
where $\Upsilon$ represents the sum of terms irrelevant to the proof. $\lambda _k$, $\mu _{k\bar k}$, $\upsilon _{\bar k}$ and $\varpi _k$ are the Lagrangian multiplier, and ${\bf{Y}}_k \in {\mathbb{C}^{N \times N}}$ is the Lagrangian multiplier matrix. The dual problem of the problem is
\begin{equation}
	\mathop {\max }\limits_{{\lambda _k},{\mu _{k\bar k}},{\upsilon _{\bar k},{\varpi _k}} > 0,{{\bf{Y}}_k} \succeq 0} {\rm{  }}\mathop {{\rm{min}}}\limits_{{{\bf{W}}_k},{\varphi _k},{\phi _{\bar k}}} {\rm{ ~~~~~}}{\cal L}.
\end{equation}
Next, we use the Karush-Kuhn-Tucker (KKT) conditions to investigate the optimal solution structure of the dual problem. The KKT condition related to ${\bf{W}}_k^ * $ can be given
\begin{equation}
	{K_1}:\lambda _k^ * ,\mu _{k\bar k}^ * ,\upsilon _{\bar k}^ *,\varpi _k  > 0,{\bf{Y}}_k^ *  \succeq 0,{K_2}:{\bf{W}}_k^ * {\bf{Y}}_k^ *  = 0,{K_3}:{\nabla _{{\bf{W}}_k^ * }}{\cal L} = 0,
\end{equation}
where $\lambda _k^ * ,\mu _{k\bar k}^ * ,\upsilon _{\bar k}^ *,\varpi _k $ and ${\bf{Y}}_k^ *$ represent the optimal Lagrangian multiplers of the dual problem. ${\nabla _{{\bf{W}}_k^ * }}{\cal L}$ denotes the gradient vector of Eq. (45) w.r.t ${\bf{W}}_k^ *$. We can explicitly express K3 as follows
\begin{equation}
	{\bf{Y}}_k^ *  = {{\bf{I}}_N} - {\bf{\Delta }},
\end{equation}
where $\bf{\Delta }$ can be given by
\begin{equation}
	{\bf{\Delta }} =   \lambda _k^ * {{\bf{H}}_k} - \sum\limits_{s\left( {\bar k} \right) > s\left( k \right)} {\mu _{k\bar k}^ * } {\rm{Tr}}\left( {\frac{1}{{{\bf{h}}_k^H{\bf{W}}_k^{\left( r \right)}{{\bf{h}}_k}}}{{\bf{H}}_k}} \right) + \sum\limits_{\bar k = 1}^{K - 1} {\upsilon _{\bar k}^ * } {{\bf{H}}_{\bar k}} + \sum\limits_{k = 1}^K {\varpi _k^ * } {{\bf{H}}_k}.
\end{equation}
Next, we will prove that the beamforming matrix ${\bf{W}}_k^ *$ is rank-one by exploring the structure of ${\bf{Y}}_k^ *$. We set the maximum eigenvalue of $\bf{\Delta }$ to ${\zeta _{\max }}$. It is worth noting that due to the randomness of the channel, the probability that multiple eigenvalues have the same maximum value is zero. According to Eq. (18), if ${\zeta _{\max }} > 1$, ${\bf{Y}}_k^ *$ cannot be positive semidefinite, which contradicts $K_1$. If ${\zeta _{\max }} < 1$, ${\bf{Y}}_k^ *$ must be positive definite and full rank. It can be seen from $K_2$ that ${\bf{W}}_k^ *$ can only be ${\bf{0}}$, which is obviously contradictory to reality. Therefore, ${\zeta _{\max }} < 1$ must hold, then ${\rm{Rank}}\left( {{\bf{Y}}_k^*} \right) = N - 1$. Therefore, ${\rm{Rank}}\left( {{\bf{W}}_k^*} \right) = 1$, i.e., the beamforming matrix ${\bf{W}}_k^*$ is rank-one. The proof is completed.
\section{Proof of Eq. (30)}
Let ${B_{kk}} = {\left| {{\bf{h}}_k^H{{\bf{w}}_k}} \right|^2}$, ${B_{jk}} = \sum\limits_{s\left( j \right) > s\left( k \right)} {{{\left| {{\bf{h}}_k^H{{\bf{w}}_j}} \right|}^2}}  + \sigma _k^2$, ${B_{\bar kk}} = {\left| {{\bf{h}}_{\bar k}^H{{\bf{w}}_k}} \right|^2}$ and ${B_{j\bar k}} = \sum\limits_{s\left( j \right) > s\left( k \right)} {{{\left| {{\bf{h}}_{\bar k}^H{{\bf{w}}_j}} \right|}^2}}  + \sigma _{\bar k}^2$. Then constraint (29c) can be rewritten as follows
\begin{equation}
	\frac{{{B_{kk}}}}{{{B_{jk}} + \frac{{\delta _k^2}}{{{\rho _k}}}}} \le \frac{{{B_{\bar kk}}}}{{{B_{j\bar k}} + \frac{{\delta _{\bar k}^2}}{{{\rho _{\bar k}}}}}},~{\rm{if}}~{\rm{ }}s\left( k \right) < s\left( {\bar k} \right),
\end{equation}
which can also be expressed as
\begin{equation}
	{B_{kk}}{B_{j\bar k}} + {B_{kk}}\delta _{\bar k}^2\frac{1}{{{\rho _{\bar k}}}} -{B_{\bar kk}}{B_{jk}} - {B_{\bar kk}}\delta _k^2\frac{1}{{{\rho _k}}}\le 0,~{\rm{if}}~{\rm{ }}s\left( k \right) < s\left( {\bar k} \right).
\end{equation}
Since the fourth term of LHS is concave, we can use SCA to obtain its upper bound, which can be given by
\begin{equation}
	 - \frac{1}{{{\rho _k}}} \le  - \frac{1}{{\rho _k^{\left( r \right)}}} + \frac{1}{{{{\left( {\rho _k^{\left( r \right)}} \right)}^2}}}\left( {{\rho _k} - \rho _k^{\left( r \right)}} \right).
\end{equation}
Therefore, constraint (29c) can be transformed into as follows Eq. (30). The proof is completed.
\section{Proof of Eq. (34)}
Let ${C_k} = \sigma _k^2 + \frac{{\delta _k^2}}{{{\rho _k}}}$ and ${C_{\bar k}} = \sigma _{\bar k}^2 + \frac{{\delta _{\bar k}^2}}{{{\rho _{\bar k}}}}$. We take the logarithm of both sides of the constraint (33c), which can be expressed as
\begin{equation}
	\ln \left( {\frac{{{\rm{Tr}}\left( {{{\bf{S}}_{k,k}}{\bf{\bar U}}} \right) + {{\left| {{{\bf{q}}_{k,k}}} \right|}^2}}}{{\sum\limits_{s\left( j \right) > s\left( k \right)} {\left( {{\rm{Tr}}\left( {{{\bf{S}}_{j,k}}{\bf{\bar U}}} \right) + {{\left| {{{\bf{q}}_{j,k}}} \right|}^2}} \right) + {C_k}} }}} \right) \le \ln \left( {\frac{{{\rm{Tr}}\left( {{{\bf{S}}_{k,\bar k}}{\bf{\bar U}}} \right) + {{\left| {{{\bf{q}}_{k,\bar k}}} \right|}^2}}}{{\sum\limits_{s\left( j \right) > s\left( k \right)} {\left( {{\rm{Tr}}\left( {{{\bf{S}}_{j,\bar k}}{\bf{\bar U}}} \right) + {{\left| {{{\bf{q}}_{j,\bar k}}} \right|}^2}} \right) + {C_{\bar k}}} }}} \right).
\end{equation}
Eq. (53) can also be given by
\begin{equation}
	\begin{aligned}
		&\ln \left( {{\rm{Tr}}\left( {{{\bf{S}}_{k,k}}{\bf{\bar U}}} \right) + {{\left| {{{\bf{q}}_{k,k}}} \right|}^2}} \right) - \ln \left( {\sum\limits_{s\left( j \right) > s\left( k \right)} {\left( {{\rm{Tr}}\left( {{{\bf{S}}_{j,k}}{\bf{\bar U}}} \right) + {{\left| {{{\bf{q}}_{j,k}}} \right|}^2}} \right) + {C_k}} } \right) \\
		& - \ln \left( {{\rm{Tr}}\left( {{{\bf{S}}_{k,\bar k}}{\bf{\bar U}}} \right) + {{\left| {{{\bf{q}}_{k,\bar k}}} \right|}^2}} \right) + \ln \left( {\sum\limits_{s\left( j \right) > s\left( k \right)} {\left( {{\rm{Tr}}\left( {{{\bf{S}}_{j,\bar k}}{\bf{\bar U}}} \right) + {{\left| {{{\bf{q}}_{j,\bar k}}} \right|}^2}} \right) + {C_{\bar k}}} } \right) \le 0.
	\end{aligned}
\end{equation}
Since the first term and the fourth term on the LHS are both concave, we apply SCA to obtain their upper bound, which can be expressed as 
\begin{equation}
	\begin{aligned}
		&\ln \left( {{\rm{Tr}}\left( {{{\bf{S}}_{k,k}}{\bf{\bar U}}} \right) + {{\left| {{{\bf{q}}_{k,k}}} \right|}^2}} \right) \le \ln \left( {{\rm{Tr}}\left( {{{\bf{S}}_{k,k}}{{{\bf{\bar U}}}^{\left( r \right)}}} \right) + {{\left| {{{\bf{q}}_{k,k}}} \right|}^2}} \right) + \\
		&{\rm{Tr}}\left( {{{\left( {\frac{1}{{{\rm{Tr}}\left( {{{\bf{S}}_{k,k}}{{{\bf{\bar U}}}^{\left( r \right)}}} \right) + {{\left| {{{\bf{q}}_{k,k}}} \right|}^2}}}{\bf{S}}_{k,k}^H} \right)}^H}\left( {{\bf{\bar U}} - {{{\bf{\bar U}}}^{\left( r \right)}}} \right)} \right) \buildrel \Delta \over = {\hat g_1}\left( {{\bf{\bar U}}} \right),
	\end{aligned}
\end{equation} 
and
\begin{equation}
	\begin{aligned}
		&\ln \left( {\sum\limits_{s\left( j \right) > s\left( k \right)} {\left( {{\rm{Tr}}\left( {{{\bf{S}}_{j,\bar k}}{\bf{\bar U}}} \right) + {{\left| {{{\bf{q}}_{j,\bar k}}} \right|}^2}} \right) + {C_{\bar k}}} } \right) \le \ln \left( {\sum\limits_{s\left( j \right) > s\left( k \right)} {\left( {{\rm{Tr}}\left( {{{\bf{S}}_{j,\bar k}}{{{\bf{\bar U}}}^{\left( r \right)}}} \right) + {{\left| {{{\bf{q}}_{j,\bar k}}} \right|}^2}} \right) + {C_{\bar k}}} } \right) \\
		&  + {\rm{Tr}}\left( {{{\left( {\frac{1}{{\left( {{\rm{Tr}}\left( {{{\bf{S}}_{j,\bar k}}{{{\bf{\bar U}}}^{\left( r \right)}}} \right) + {{\left| {{{\bf{q}}_{j,\bar k}}} \right|}^2}} \right) + {C_{\bar k}}}}\sum\limits_{s\left( j \right) > s\left( k \right)} {{\bf{S}}_{j,\bar k}^H} } \right)}^H}\left( {{\bf{\bar U}} - {{{\bf{\bar U}}}^{\left( r \right)}}} \right)} \right) \buildrel \Delta \over = {{\hat g}_2}\left( {{\bf{\bar U}}} \right).
	\end{aligned}
\end{equation} 
Therefore, constraint (33c) can be transformed into as follows Eq. (34). The proof is completed.
\ifCLASSOPTIONcaptionsoff
  \newpage
\fi



\bibliographystyle{IEEEtran}
\bibliography{reference}

\begin{thebibliography}{10}
\providecommand{\url}[1]{#1}
\csname url@samestyle\endcsname
\providecommand{\newblock}{\relax}
\providecommand{\bibinfo}[2]{#2}
\providecommand{\BIBentrySTDinterwordspacing}{\spaceskip=0pt\relax}
\providecommand{\BIBentryALTinterwordstretchfactor}{4}
\providecommand{\BIBentryALTinterwordspacing}{\spaceskip=\fontdimen2\font plus
\BIBentryALTinterwordstretchfactor\fontdimen3\font minus
  \fontdimen4\font\relax}
\providecommand{\BIBforeignlanguage}[2]{{%
\expandafter\ifx\csname l@#1\endcsname\relax
\typeout{** WARNING: IEEEtran.bst: No hyphenation pattern has been}%
\typeout{** loaded for the language `#1'. Using the pattern for}%
\typeout{** the default language instead.}%
\else
\language=\csname l@#1\endcsname
\fi
#2}}
\providecommand{\BIBdecl}{\relax}
\BIBdecl

\bibitem{8910627}
Q.~{Wu} and R.~{Zhang}, ``Towards smart and reconfigurable environment:
  Intelligent reflecting surface aided wireless network,'' \emph{IEEE Commun.
  Mag.}, vol.~58, no.~1, pp. 106--112, 2020.

\bibitem{ni2020resource}
W.~Ni, X.~Liu, Y.~Liu, H.~Tian, and Y.~Chen, ``Resource allocation for
  multi-cell {IRS}-aided {NOMA} networks,'' \emph{arXiv preprint
  arXiv:2006.11811}, 2020.

\bibitem{6798744}
L.~{Lu}, G.~Y. {Li}, A.~L. {Swindlehurst}, A.~{Ashikhmin}, and R.~{Zhang}, ``An
  overview of massive {MIMO}: Benefits and challenges,'' \emph{IEEE J. Sel.
  Topics Signal Process.}, vol.~8, no.~5, pp. 742--758, 2014.

\bibitem{6995966}
Q.~{Wu}, W.~{Chen}, M.~{Tao}, J.~{Li}, H.~{Tang}, and J.~{Wu}, ``Resource
  allocation for joint transmitter and receiver energy efficiency maximization
  in downlink {OFDMA} systems,'' \emph{IEEE Trans. Commun.}, vol.~63, no.~2,
  pp. 416--430, 2015.

\bibitem{liu2017non}
Y.~Liu, Z.~Qin, M.~Elkashlan, Z.~Ding, A.~Nallanathan, and L.~Hanzo,
  ``Non-orthogonal multiple access for {5G} and beyond,'' \emph{Proc. IEEE},
  vol. 105, no.~12, pp. 2347--2381, 2017.

\bibitem{7263349}
L.~{Dai}, B.~{Wang}, Y.~{Yuan}, S.~{Han}, I.~{Chih-lin}, and Z.~{Wang},
  ``Non-orthogonal multiple access for 5g: solutions, challenges,
  opportunities, and future research trends,'' \emph{IEEE Commun. Mag.},
  vol.~53, no.~9, pp. 74--81, 2015.

\bibitem{8493528}
Y.~{Zhou}, V.~W.~S. {Wong}, and R.~{Schober}, ``Coverage and rate analysis of
  millimeter wave {NOMA} networks with beam misalignment,'' \emph{IEEE Trans.
  Wireless Commun.}, vol.~17, no.~12, pp. 8211--8227, 2018.

\bibitem{7752784}
F.~{Wei} and W.~{Chen}, ``Low complexity iterative receiver design for sparse
  code multiple access,'' \emph{IEEE Trans. Commun.}, vol.~65, no.~2, pp.
  621--634, 2017.

\bibitem{8674826}
Y.~Feng, S.~Yan, Z.~Yang, N.~Yang, and J.~Yuan, ``Beamforming design and power
  allocation for secure transmission with {NOMA},'' \emph{IEEE Trans. Wireless
  Commun.}, vol.~18, no.~5, pp. 2639--2651, 2019.

\bibitem{8085125}
Y.~{Cai}, Z.~{Qin}, F.~{Cui}, G.~Y. {Li}, and J.~A. {McCann}, ``Modulation and
  multiple access for {5G} networks,'' \emph{IEEE Communications Surveys
  Tutorials}, vol.~20, no.~1, pp. 629--646, 2018.

\bibitem{7842433}
Z.~{Ding}, Y.~{Liu}, J.~{Choi}, Q.~{Sun}, M.~{Elkashlan}, I.~{Chih-Lin}, and
  H.~V. {Poor}, ``Application of non-orthogonal multiple access in {LTE} and
  {5G} networks,'' \emph{IEEE Commun. Mag.}, vol.~55, no.~2, pp. 185--191,
  2017.

\bibitem{7273963}
Z.~{Ding}, P.~{Fan}, and H.~V. {Poor}, ``Impact of user pairing on {5G}
  nonorthogonal multiple-access downlink transmissions,'' \emph{IEEE Trans.
  Veh. Technol.}, vol.~65, no.~8, pp. 6010--6023, 2016.

\bibitem{7867826}
Y.~{Zeng}, B.~{Clerckx}, and R.~{Zhang}, ``Communications and signals design
  for wireless power transmission,'' \emph{IEEE Trans. Commun.}, vol.~65,
  no.~5, pp. 2264--2290, 2017.

\bibitem{8476597}
B.~{Clerckx}, R.~{Zhang}, R.~{Schober}, D.~W.~K. {Ng}, D.~I. {Kim}, and H.~V.
  {Poor}, ``Fundamentals of wireless information and power transfer: From {RF}
  energy harvester models to signal and system designs,'' \emph{IEEE J. Sel.
  Areas Commun.}, vol.~37, no.~1, pp. 4--33, 2019.

\bibitem{6805330}
Q.~{Shi}, L.~{Liu}, W.~{Xu}, and R.~{Zhang}, ``Joint transmit beamforming and
  receive power splitting for {MISO} {SWIPT} systems,'' \emph{IEEE Trans.
  Wireless Commun.}, vol.~13, no.~6, pp. 3269--3280, 2014.

\bibitem{6623062}
X.~{Zhou}, R.~{Zhang}, and C.~K. {Ho}, ``Wireless information and power
  transfer: Architecture design and rate-energy tradeoff,'' \emph{IEEE Trans.
  Commun.}, vol.~61, no.~11, pp. 4754--4767, 2013.

\bibitem{8811733}
Q.~{Wu} and R.~{Zhang}, ``Intelligent reflecting surface enhanced wireless
  network via joint active and passive beamforming,'' \emph{IEEE Trans.
  Wireless Commun.}, vol.~18, no.~11, pp. 5394--5409, 2019.

\bibitem{8930608}
Q.~{Wu} and R.~{Zhang}, ``Beamforming optimization for wireless network aided
  by intelligent reflecting surface with discrete phase shifts,'' \emph{IEEE
  Trans. Commun.}, vol.~68, no.~3, pp. 1838--1851, 2020.

\bibitem{9048622}
W.~{Tang}, M.~Z. {Chen}, J.~Y. {Dai}, Y.~{Zeng}, X.~{Zhao}, S.~{Jin},
  Q.~{Cheng}, and T.~J. {Cui}, ``Wireless communications with programmable
  metasurface: New paradigms, opportunities, and challenges on transceiver
  design,'' \emph{IEEE Wireless Commun.}, vol.~27, no.~2, pp. 180--187, 2020.

\bibitem{9136592}
C.~{Huang}, S.~{Hu}, G.~C. {Alexandropoulos}, A.~{Zappone}, C.~{Yuen},
  R.~{Zhang}, M.~D. {Renzo}, and M.~{Debbah}, ``Holographic {MIMO} surfaces for
  {6G} wireless networks: Opportunities, challenges, and trends,'' \emph{IEEE
  Wireless Commun.}, vol.~27, no.~5, pp. 118--125, 2020.

\bibitem{yuan2020reconfigurable}
X.~Yuan, Y.-J. Zhang, Y.~Shi, W.~Yan, and H.~Liu,
  ``Reconfigurable-intelligent-surface empowered {6G} wireless communications:
  Challenges and opportunities,'' \emph{arXiv preprint arXiv:2001.00364}, 2020.

\bibitem{8466374}
C.~{Liaskos}, S.~{Nie}, A.~{Tsioliaridou}, A.~{Pitsillides}, S.~{Ioannidis},
  and I.~{Akyildiz}, ``A new wireless communication paradigm through
  software-controlled metasurfaces,'' \emph{IEEE Commun. Mag.}, vol.~56, no.~9,
  pp. 162--169, 2018.

\bibitem{rajatheva2020white}
N.~Rajatheva, I.~Atzeni, E.~Bjornson, A.~Bourdoux, S.~Buzzi, J.-B. Dore,
  S.~Erkucuk, M.~Fuentes, K.~Guan, Y.~Hu \emph{et~al.}, ``White paper on
  broadband connectivity in {6G},'' \emph{arXiv preprint arXiv:2004.14247},
  2020.

\bibitem{gong2019towards}
S.~Gong, X.~Lu, D.~T. Hoang, D.~Niyato, L.~Shu, D.~I. Kim, and Y.-C. Liang,
  ``Towards smart radio environment for wireless communications via intelligent
  reflecting surfaces: A comprehensive survey,'' \emph{arXiv preprint
  arXiv:1912.07794}, 2019.

\bibitem{zhao2019survey}
J.~Zhao, ``A survey of intelligent reflecting surfaces {(IRSs)}: Towards {6G}
  wireless communication networks with massive {MIMO} 2.0,'' \emph{arXiv
  preprint arXiv:1907.04789}, 2019.

\bibitem{9117136}
W.~{Yan}, X.~{Yuan}, Z.~Q. {He}, and X.~{Kuai}, ``Passive beamforming and
  information transfer design for reconfigurable intelligent surfaces aided
  multiuser {MIMO} systems,'' \emph{IEEE J. Sel. Areas Commun.}, vol.~38,
  no.~8, pp. 1793--1808, 2020.

\bibitem{9159923}
L.~{Dong} and H.~{Wang}, ``Enhancing secure {MIMO} transmission via intelligent
  reflecting surface,'' \emph{IEEE Trans. Wireless Commun.}, pp. 1--1, 2020.

\bibitem{9201173}
S.~{Hong}, C.~{Pan}, H.~{Ren}, K.~{Wang}, and A.~{Nallanathan},
  ``Artificial-noise-aided secure {MIMO} wireless communications via
  intelligent reflecting surface,'' \emph{IEEE Trans. Commun.}, pp. 1--1, 2020.

\bibitem{9355404}
K.~Zhi, C.~Pan, H.~Ren, and K.~Wang, ``Statistical {CSI}-based design for
  reconfigurable intelligent surface-aided massive {MIMO} systems with direct
  links,'' \emph{IEEE Wireless Commun. Lett.}, pp. 1--1, 2021.

\bibitem{9133107}
T.~{Bai}, C.~{Pan}, Y.~{Deng}, M.~{Elkashlan}, A.~{Nallanathan}, and
  L.~{Hanzo}, ``Latency minimization for intelligent reflecting surface aided
  mobile edge computing,'' \emph{IEEE J. Sel. Areas Commun.}, vol.~38, no.~11,
  pp. 2666--2682, 2020.

\bibitem{9234511}
T.~{Shafique}, H.~{Tabassum}, and E.~{Hossain}, ``Optimization of wireless
  relaying with flexible {UAV}-borne reflecting surfaces,'' \emph{IEEE Trans.
  Commun.}, pp. 1--1, 2020.

\bibitem{8959174}
S.~{Li}, B.~{Duo}, X.~{Yuan}, Y.~{Liang}, and M.~{Di Renzo}, ``Reconfigurable
  intelligent surface assisted {UAV} communication: Joint trajectory design and
  passive beamforming,'' \emph{IEEE Wireless Commun. Lett.}, vol.~9, no.~5, pp.
  716--720, 2020.

\bibitem{8743496}
H.~{Shen}, W.~{Xu}, S.~{Gong}, Z.~{He}, and C.~{Zhao}, ``Secrecy rate
  maximization for intelligent reflecting surface assisted {Multi}-antenna
  communications,'' \emph{IEEE Commun. Lett.}, vol.~23, no.~9, pp. 1488--1492,
  2019.

\bibitem{9248012}
L.~{Lv}, Q.~{Wu}, Z.~{Li}, N.~{Al-Dhahir}, and J.~{Chen}, ``Secure two-way
  communications via intelligent reflecting surfaces,'' \emph{IEEE Commun.
  Lett.}, pp. 1--1, 2020.

\bibitem{9390203}
C.~Wu, S.~Yan, X.~Zhou, R.~Chen, and J.~Sun, ``Intelligent reflecting surface
  ({IRS})-aided covert communication with {Warden}'s statistical {CSI},''
  \emph{IEEE Wireless Commun. Lett.}, pp. 1--1, 2021.

\bibitem{9180053}
G.~Zhou, C.~Pan, H.~Ren, K.~Wang, and A.~Nallanathan, ``A framework of robust
  transmission design for {IRS}-aided {MISO} communications with imperfect
  cascaded channels,'' \emph{IEEE Trans. Signal Process.}, vol.~68, pp.
  5092--5106, 2020.

\bibitem{fu2019reconfigurable}
M.~Fu, Y.~Zhou, Y.~Shi, and K.~B. Letaief, ``Reconfigurable intelligent surface
  empowered downlink non-orthogonal multiple access,'' \emph{arXiv preprint
  arXiv:1910.07361}, 2019.

\bibitem{8970580}
B.~{Zheng}, Q.~{Wu}, and R.~{Zhang}, ``Intelligent reflecting surface-assisted
  multiple access with user pairing: {NOMA} or {OMA}?'' \emph{IEEE Commun.
  Lett.}, vol.~24, no.~4, pp. 753--757, 2020.

\bibitem{9167258}
J.~{Zuo}, Y.~{Liu}, Z.~{Qin}, and N.~{Al-Dhahir}, ``Resource allocation in
  intelligent reflecting surface assisted {NOMA} systems,'' \emph{IEEE Trans.
  Commun.}, pp. 1--1, 2020.

\bibitem{9203956}
M.~{Zeng}, X.~{Li}, G.~{Li}, W.~{Hao}, and O.~A. {Dobre}, ``Sum rate
  maximization for {IRS}-assisted uplink {NOMA},'' \emph{IEEE Commun. Lett.},
  pp. 1--1, 2020.

\bibitem{9140006}
J.~{Zuo}, Y.~{Liu}, E.~{Basar}, and O.~A. {Dobre}, ``Intelligent reflecting
  surface enhanced millimeter-wave {NOMA} systems,'' \emph{IEEE Commun. Lett.},
  pp. 1--1, 2020.

\bibitem{9240028}
J.~{Zhu}, Y.~{Huang}, J.~{Wang}, K.~{Navaie}, and Z.~{Ding}, ``Power efficient
  {IRS}-assisted {NOMA},'' \emph{IEEE Trans. Commun.}, pp. 1--1, 2020.

\bibitem{8941080}
Q.~{Wu} and R.~{Zhang}, ``Weighted sum power maximization for intelligent
  reflecting surface aided {SWIPT},'' \emph{IEEE Wireless Commun. Lett.},
  vol.~9, no.~5, pp. 586--590, 2020.

\bibitem{9133435}
Q.~{Wu} and R.~{Zhang}, ``Joint active and passive beamforming optimization for
  intelligent reflecting surface assisted {SWIPT} under {QoS} constraints,''
  \emph{IEEE J. Sel. Areas Commun.}, vol.~38, no.~8, pp. 1735--1748, 2020.

\bibitem{9110849}
C.~{Pan}, H.~{Ren}, K.~{Wang}, M.~{Elkashlan}, A.~{Nallanathan}, J.~{Wang}, and
  L.~{Hanzo}, ``Intelligent reflecting surface aided {MIMO} broadcasting for
  simultaneous wireless information and power transfer,'' \emph{IEEE J. Sel.
  Areas Commun.}, vol.~38, no.~8, pp. 1719--1734, 2020.

\bibitem{6231145}
L.~{Subrt} and P.~{Pechac}, ``Intelligent walls as autonomous parts of smart
  indoor environments,'' \emph{IET Commun.}, vol.~6, no.~8, pp. 1004--1010,
  2012.

\bibitem{8170332}
J.~{Cui}, Y.~{Liu}, Z.~{Ding}, P.~{Fan}, and A.~{Nallanathan}, ``Optimal user
  scheduling and power allocation for millimeter wave {NOMA} systems,''
  \emph{IEEE Trans. Wireless Commun.}, vol.~17, no.~3, pp. 1502--1517, 2018.

\bibitem{grant2014cvx}
M.~Grant and S.~Boyd, ``{CVX}: Matlab software for disciplined convex
  programming, version 2.1,'' 2014.

\bibitem{ben2001lectures}
A.~Ben-Tal and A.~Nemirovski, \emph{Lectures on modern convex optimization:
  analysis, algorithms, and engineering applications}.\hskip 1em plus 0.5em
  minus 0.4em\relax SIAM, 2001.

\bibitem{boyd2004convex}
S.~Boyd, S.~P. Boyd, and L.~Vandenberghe, \emph{Convex optimization}.\hskip 1em
  plus 0.5em minus 0.4em\relax Cambridge university press, 2004.

\end{thebibliography}
\end{document}